\documentstyle[12pt,graphicx]{article}
\title{{ Naturalness, Weak Scale Supersymmetry and 
the Prospect
for  the Observation of Supersymmetry at the Tevatron and at the LHC}}
\author{Kwok Lung Chan, Utpal Chattopadhyay 
  and Pran Nath\\
Department of Physics, Northeastern University \\
Boston, MA 02115-5005}
\date{}
\begin{document}
\maketitle  
\abstract{ 
Naturalness bounds on weak scale supersymmetry in the context of
radiative breaking of the electro-weak symmetry are analyzed. 
In the case of minimal supergravity it is found that for low tan$\beta$
and for low values of fine tuning $\Phi$, where $\Phi$ is defined essentially 
by the
ratio $\mu^2/M_Z^2$ where $\mu$ is the Higgs mixing parameter and $M_Z$ 
is the Z boson mass, the allowed values of the universal
scalar parameter $m_0$, and the universal gaugino mass $m_{1/2}$ lie on the
surface of an ellipsoid with radii fixed by $\Phi$ leading to 
tightly constrained  
upper bounds $\sim \sqrt\Phi$.
Thus for $tan\beta\leq 2(\leq 5)$ it is found that the upper limits for the
entire set of 
sparticle masses lie in the range $<$ 700 GeV $(<1.5$TeV) for any
reasonable range of fine tuning ($\Phi\leq 20$). However, it is found that
there exist regions of the parameter space where the fine tuning 
does not tightly constrain $m_0$ and $m_{1/2}$. 
 Effects of non-universalities in the Higgs
sector and in the third generation sector on 
naturalness bounds are also analyzed and it is found that non-universalities
can significantly affect the upper bounds.
 It is also found that achieving the maximum
Higgs mass allowed in supergravity unified models requires a high degree 
of fine tuning. Thus a heavy sparticle spectrum is indicated if the Higgs
mass exceeds 120 GeV. 
The prospect for the discovery of supersymmetry
at the Tevatron and at the LHC in view of these results 
is  discussed.}

\def\til#1{#1 \over {4 \pi}}
\def\Mz2{M_Z^2}
\def\mh{m_{1/2}}
\def\sms#1#2{m^2_{\tilde#1,#2}}
\newtheorem{theorem}{Theorem}
\section{Introduction}

One of the important elements in supersymmetric model building
is the  issue of the mass scale of the supersymmetric particles. 
There is the general expectation that this scale should be of 
the order of the scale of the electro-weak physics, i.e., in the range
of a TeV.  This idea is given a more concrete meaning in the context
of supergravity unification\cite{applied} where one has  spontaneous breaking
of the  electro-weak symmetry by radiative corrections\cite{inoue}. Radiative 
breaking
of the electro-weak symmetry relates the scale of supersymmetry soft breaking
terms  directly to the Z boson mass. This relationship then tells us
 that the soft SUSY breaking scale should not be much larger than the
scale of the Z boson mass otherwise a significant fine tuning will be 
needed to recover the Z boson mass. The above general connection would be 
thwarted if there were large internal cancellations occurring naturally
within the radiative breaking condition which would allow $m_0$ and 
$m_{1/2}$ disproportionately large for a fixed fine tuning. We shall show
that precisely such a situation does arise in certain domains of the
supergravity parameter space.

  The simplest fine tuning criterion is to impose the constraint that 
  $m_0, m_{\tilde g}<1$ TeV where $m_0$ is the universal soft SUSY
  breaking scalar mass in minimal supergravity and $m_{\tilde g}$
 is the gluino mass. The above criterion is easy  to implement and has 
 been used widely in the literature (for a review see Ref.\cite{swieca}).    
	 A  more involved fine tuning criterion 
    is given  in Ref.\cite{BG}. 
 However, it appears that the criterion of Ref.\cite{BG} 
  is actually a measure of the sensitivity rather than of fine tuning\cite{carlos,AC}.  
Another naturalness criterion is proposed in Ref.\cite{AC} and 
involves a distribution function. 
Although the distribution function is  arbitrary 
the authors  show that different choices of the  function lead numerically
to similar fine tuning limits.

In the analysis of this paper we use the fine tuning criterion introduced in
Ref.\cite{aosta} in terms of the Higgs mixing parameter $\mu$
 which has several 
attractive features. It is 
 physically well motivated, free of ambiguities and 
 easy to implement. Next we use the criterion to  analyze the 
upper limits of sparticle masses for low  values of
tan$\beta$, i.e., tan$\beta\leq 5$.
In this case one finds that $m_0$ and $m_{1/2}$ allowed by radiative breaking
lie on the surface of an ellipsoid, and hence the upper limits of the 
sparticle masses are directly controlled by the radii of the ellipsoid 
which in turn are determined by the choice of fine tuning.  
  For instance, one finds that  if one is in the low tan$\beta$ end of 
  $b-\tau$ unification\cite{barger}
   with the top mass in the experimental range, i.e.
    tan$\beta\approx 2$, then 
    for any reasonable range of fine tuning  
the sparticle mass upper limits for the entire set of SUSY particles
lie within the mass range below 1 TeV. Further, one finds that the 
light Higgs mass lies below  90 GeV under the same constraints.
 Thus in this case discovery of
supersymmetry at the LHC is guaranteed according to any reasonable fine
tuning  criterion. 
	Next the paper explores larger values of tan$\beta$, i.e.,
	tan$\beta\geq 10$ and here one finds that $m_0$ and $m_{1/2}$
	for moderate values of fine tuning do not lie on the surface 
	of an ellipsoid; rather one finds that they lie on the surface of 
	a hyperboloid. In this case $m_0$ and $m_{1/2}$ are not 
	bounded by the $\mu$ constraint equation and large values
	of $m_0$ and $m_{1/2}$ can result with a fixed fine tuning.

	Effect of non-universalities on naturalness  is also analyzed.
	 Again one finds phenomena similar to the ones discussed 
	 above, although the domains in which these  phenomena occur
	 are  shifted relative to those in the universal case.  
	One of the important results that emerges is that  the upper limits
of sparticle masses can be dramatically affected  by non-universalities. 
These results have important implications for the discovery of 
supersymmetry at the Tevatron and the LHC.

Our analysis is carried out in the framework of supergravity 
models with gravity mediated breaking of 
supersymmetry\cite{chams,applied,swieca}. This class of models possesses
many attractive features. One of the more attractive features of these 
models is that with R parity invariance the lightest neutralino is 
also the lightest supersymmetric particle over most of the parameter space
of the theory 
and hence a candidate for
cold dark matter. Precision renormalization 
group analyses  show\cite{ross} that these models can accommodate just the right
amount of  dark matter consistent with the current astrophysical 
data\cite{dark,barger2}. 
However, in this work we shall not impose the constraint
of dark matter.

The outline of the paper is as follows: In Sec.2 we 
give a brief discussion of the fine tuning measure used in the 
analysis. In Sec.3 we use this criterion to discuss the upper
limits on the sparticle masses in minimal supergravity for low
tan$\beta$, i.e., tan$\beta\leq 5$ and show that the allowed solutions
to radiative breaking lie on the surface of  an ellipsoid.
In Sec.4 we discuss  naturalness in beyond the low tan$\beta$ region.
  Here we show that radiative breaking of the
electro-weak symmetry leads to the soft SUSY breaking parameters lying
on the surface of a hyperboloid. In Sec.5 we discuss the effects of
non-universalities on the upper limits.  
In Sec.6  we show that a high degree of
fine tuning is needed to have the light Higgs mass approach its maximum upper 
limit. The limits on $\Phi$ from the current data are discussed in
sec.7. 
Implications of these results for the discovery
of supersymmetric particles at colliders is also discussed in Secs. 3-6.
  Conclusions are given in Sec.8.

\section{ Measure of Naturalness}

\def\tana2{\tan^2\alpha_H}
\def\tanb2{\tan^2\beta}
\def\lamp{\lambda_+}
\def\lamm{\lambda_-}
\def\lampm{\lambda_\pm}
\def\m0pr{m_0'}
\def\mhpr{m_{1/2}'}
\def\massmat{
	\left[\begin{array}{cc}
        p       &       q       \\
        q       &       r
	\end{array}\right]}
\def\massvec{\left[\begin{array}{cc}
        	m_0 & m_{1/2}
		\end{array}\right]}
\def\masscol{\left[\begin{array}{c}
        m_0     \\
        m_{1/2} \end{array}\right]}
We give below an improved version of the   
analysis of the fine tuning criterion given in Ref.\cite{aosta}. 
The radiative electro-weak symmetry breaking condition is given by 

\begin{equation} 
\frac{1}{2}M_Z^2=\lambda^2- \mu^2
\label{b1 } 
\end{equation}
where $\lambda^2$ is defined by 
\begin{equation} 
\lambda^2   =  \frac{\bar m_{H_1}^2 - \bar m_{H_2}^2 
\tan^2\beta}{\tan^2\beta - 1}\label{ } 
\end{equation}
Here $\bar m_{H_i}^2=m_{H_i}^2+\Sigma_i$(i=1,2) where 
$\Sigma_i$ arise from the one  loop corrections to the effective 
potential\cite{gamberini}.
 The issue of fine tuning
now revolves around the fact that  a cancellation is needed between the
$\lambda^2$ term and the $\mu^2$ term to arrange the correct experimental 
value of $M_Z$. Thus a large value of $\lambda^2$
would require  a large  cancellation 
from the $\mu^2$ term resulting in
a large fine tuning.  This idea can be quantified by  defining 
the fine tuning parameter $\Phi$ so that
\begin{equation} 
\Phi^{-1}=4 \frac{\lambda^2-\mu^2}{\lambda^2+\mu^2}\label{b3 } 
\end{equation}
(The factor of 4 on the right hand side in Eq.(3) is just a convenient 
normalization.)
 The expression for $\Phi$ can be simplified by inserting in the radiative 
breaking condition Eq.(1). We then get
\begin{equation} 
\Phi=\frac{1}{4}+\frac{\mu^2}{M_Z^2}\label{b4 } 
\end{equation}
The result above is valid with the inclusion of both the tree and the 
loop corrections
to the effective potential.($\Phi$ is related to the fine tuning parameter
$\delta$ defined in Ref.\cite{aosta} by $\Phi =\delta^{-1}$).
 For large $\mu$ one has $\Phi\sim \frac{\mu^2}{M_Z^2}$, a 
 result which  has a very direct intuitive meaning.
 A large $\mu$ implies a large cancellation 
 between the $\lambda^2$ term and
the $\mu^2$ term in Eq.1 to recover the Z boson mass and thus leads to 
 a large fine tuning. 
   Typically a large $\mu$ implies large values  for the  soft supersymmetry 
 breaking
parameters $m_0$ and $m_{1/2}$  and thus large values for the sparticle  masses.
 However, large cancellation can be enforced by the internal dynamics of 
 radiative breaking itself. In this case a small $\mu$ and hence 
 a small fine tuning allows for relatively large values of $m_0$ 
 and of $m_{1/2}$. We show that precisely such a situation arises
 for certain regions of the parameter space of both the minimal model as
 well as for models with non-universalities. 

\section{Upper Bounds on Sparticle Masses in Minimal Supergravity}

We discuss now the upper bounds  on the sparticle masses that arise 
under the criterion of fine tuning we have discussed above. 
Using the radiative electro-weak symmetry breaking constraint and ignoring
the b-quark couplings, justified for small tan$\beta$, we may express
the fine tuning parameter $\Phi_0 $ in the form 
\begin{equation}
 \Phi_0=-\frac{1}{4}+(\frac{m_0}{M_Z})^2 C_1+(\frac{A_0}{M_Z})^2 C_2 +
 (\frac{m_{\frac{1}{2}}}{M_Z})^2C_3+(\frac{m_{\frac{1}{2}}
A_0}{M_Z^2})C_4+\frac{\Delta \mu^2_{loop}}{M_Z^2}
\end{equation}
where 
\begin{equation}
C_1=\frac{1}{t^2-1}(1-\frac{3 D_0-1}{2}t^2), C_2=\frac{t^2}{t^2-1}k
\end{equation}

\begin{equation}
~C_3=\frac{1}{t^2-1}(g- t^2 e), 
~C_4=-\frac{t^2}{t^2-1}f,
\Delta \mu^2_{loop}=\frac{\Sigma_1- t^2\Sigma_2}{t^2-1}
\end{equation}
 Here $t\equiv tan\beta$, e,f,g,k and the sign conventions of $A_0$
 and $\mu$ are as defined in
 Ref.\cite{ilm}, $D_0$ is defined by 
 \begin{equation}
 D_0=1-(m_t/m_f)^2, ~~ m_f\simeq 200 sin\beta ~GeV
 \end{equation} 
 and $\Sigma_1$ and $\Sigma_2$ are as defined in Ref. \cite{gamberini}.

	To investigate the upper limits on $m_0$ and $m_{1/2}$ 
	consistent with a given fine tuning it is instructive to write 
	Eq.(5) in the form 
	
	\begin{equation}
	C_1m_0^2+C_3m'^2_{1/2}+C_2'A_0^2+\Delta \mu^2_{loop}=
	M_Z^2(\Phi_0+\frac{1}{4})
	\end{equation}
	where 
	\begin{equation}
	m_{1/2}'=m_{1/2}+\frac{1}{2}A_0\frac{C_4}{C_3},~~
	C_2'=C_2-\frac{1}{4}\frac{C_4^2}{C_3}
	\end{equation}
	and $\Delta \mu_{loop}^2$ is the loop correction. 
	 Now for the universal case one finds that the loop
corrections to $\mu$ are generally small for tan$\beta\leq 5$  
in the region of fine tuning of $\Phi_0\leq 20$.
	 Further, using renormalization
	group analysis one finds that $C_2'>0$ and $C_3>0$ and at least 
	for the range of fine tuning $\Phi_0 \leq 20$, $C_1>0$(see Table 1). 
	Thus in this case defining
\begin{equation}
a^2=M_Z^2\frac{\Phi+\frac{1}{4}}{C_3},
b^2=M_Z^2\frac{\Phi+\frac{1}{4}}{C_1},~~
c^2=M_Z^2\frac{\Phi+\frac{1}{4}}{C_2'}
\end{equation}
we find that for $tan\beta \leq 5$, $\Phi_0 \leq 20$ the radiative breaking 
condition can  be approximated by	
\begin{equation}
\frac{m_{1/2}'^2}{a^2}+\frac{m_0^2}{b^2}+\frac{A_0^2}{c^2}\simeq 1
\end{equation}
 and the renormalization
group analysis shows that at the scale $Q=M_Z$ the quantities
 $a^2$, $b^2$ an  $c^2$ are  positive. 
 Fixing the fine tuning parameter $\Phi_0$ fixes a,b, and c and one finds that 
  $m_0$ and $m_{1/2}$ are bounded as they lie on the boundary of an ellipse. 
 Further Eq.(12) implies that 
  the upper bounds on $m_0$ and $m_{1/2}$ increase as $\sim \sqrt{\Phi_0}$
  for large $\Phi_0$. A  similar dependence on fine tuning was observed in 
  the analysis of ref. 4. 

We give now the full analysis without the approximation of Eq.(12).
We consider the case of tan$\beta=2$ first
 which lies close to the low
end of the tan$\beta$ region of  b-$\tau$ unification 
with the top mass taken to lie in the experimental range\cite{barger}.
In Fig.1a. we give the contour plot of the upper limits for the 
parameters $m_0$ and $m_{1/2}$ in the $m_0-m_{1/2}$ plane for the
case of tan$\beta=2$  and $m_t=175$ GeV for  $2.5\leq\Phi_0\leq20$.
 As expected,  one finds that the contours corresponding to
larger values of  $m_0$ and $m_{1/2}$  require 
larger values of $\Phi_0$. The upper limits of the 
mass spectra for the same set of parameters as in Fig.1a are analyzed
in Fig.1b - Fig.1d. In Fig.1b the upper limits of the
mass spectra of the heavy Higgs, the first two generation squarks, and the
gluino are given. We  find that 
the mass of the squark and of the gluino are very similar over essentially
the entire range of $\Phi_0$. Upper limits of $\tilde e$, $\tilde t_1$,
$\tilde t_2$ are given in Fig.1c. 
In Fig.1d we exhibit the upper limits for
the light Higgs, the chargino, and  the lightest neutralino. We note
that except for small  values of $\Phi_0$  one finds that the scaling
laws \cite{an}(e.g. $m_{\chi_1^{0}} \simeq \frac{1}{2}   m_{\chi_1^{\pm}}$) are
obeyed  with a high degree of accuracy. 
We note that the Higgs mass upper limit in this case falls below 85-90 GeV
for $\Phi_0\leq 20$.
At the Tevatron in the Main Injector era one
will be able to detect charginos using the trileptonic 
signal\cite{trilep} 
with masses up to 230 GeV with 10fb$^{-1}$
of integrated luminosity\cite{kamon,amidei}. 
Reference to Fig.1d shows that the above 
implies that the upper limit of chargino masses for the full range of
 $\Phi_0\leq 20$ will be accessible at the Tevatron.

 For the gluino the
 mass range
 up to 450 GeV will be accessible at the Tevatron in the Main Injector 
 era with 25fb$^{-1}$ of integrated luminosity. This means that one 
 can explore gluino mass limits  up to $\Phi_0$=10
  for  tan$\beta=2$.
 However, at the LHC   gluino masses in the range 
 1.6-2.3 TeV \cite{baer}/1.4-2.6 TeV\cite{ATLAS} 
 for most values of $\mu$ and tan$\beta$ will be 
 accessible  and recent analyses show that the accuracy of the
 $m_{\tilde g}$ mass measurement can be quite good, i.e., to within 1-10\%
 depending  on what part of the supergravity parameter space one
 is in\cite{hinch}.
 Thus for tan$\beta=2$ one will be able to observe and measure 
 with reasonable accuracy the masses of the charginos, the gluino, 
	and the  squarks for the full range of 
 values of $\Phi_0\leq 20$ at the LHC. 
  It has recently been argued that  the NLC, where 
 even more accurate mass measurements\cite{tsukamoto,feng,snowmass}
 are possible, will allow one to use this device for the 
 exploration of physics at the post-GUT and string scales\cite{planck}.
 The NLC also offers the
 possibility of testing  a good  part of the parameter space
 for the tan$\beta=2$ model.
 The analysis given in Table 2 shows that the full sparticle mass
  spectrum for tan$\beta$=2  can be tested at the NLC with
  $\sqrt s=1$ TeV for $\Phi_0\leq$10 and  over the entire range 
  $\Phi_0\leq 20$  with   $\sqrt s=1.5$ TeV.
 
 We discuss next the upper limit of sparticle masses for 
 tan$\beta=5$. In Fig.2a we give the contour plot of 
  $m_0$ and $m_{1/2}$ upper limits  in the $m_0-m_{1/2}$ plane for 
   the same value of the top mass  
and in the same $\Phi_0$ range  as in Fig.1a. Here
we find that for fixed $\Phi_0$ the contours are
significantly further outwards compared  to the case for $tan\beta=2$.
Correspondingly the  upper limits of the 
mass spectra for the same value of  $\Phi_0$ are significantly larger in 
Fig. 2b-2d relative to those given in Figs. 1b-1d. 
In this case the light chargino mass lies below 243 GeV for 
$\Phi_0\leq 20$ and thus the upper limits for values of
$\Phi_0\leq 20$ could be probed at the Tevatron in the Main Injector era 
where chargino masses up to 280 GeV will be accessible 
with 100fb$^{-1}$ of integrated luminosity\cite{kamon,amidei}. 
Similarly in this case the gluino mass lies below 873 GeV for 
$\Phi_0\leq 20$ and thus the upper limit for values of
$\Phi_0\leq 20$ could be probed at the LHC which as mentioned above can 
probe gluino masses in the mass range of 
$1.6-2.3$ TeV\cite{baer}/ $1.4-2.6$ TeV\cite{ATLAS}.
LHC can probe squark masses up to 2-2.5 TeV, so squark masses of  
the above size  should be accessible at the LHC.  
 	A  full summary of 
 	the results for values of tan$\beta$=2-20 is given in Table 2 where 
 	the sparticle mass limits in the range $2.5\leq \Phi_0\leq 20$
 	are given. The analysis
 	tells us that for a reasonable constraint on $\Phi_0$, i.e. 
 	 $\Phi_0\leq 20$, the gluino and the squarks  must be discovered 
 	 at the LHC for the values of tan$\beta\leq 5$.

\section{Regions Of the Hyperbolic Constraint}
In this section we discuss  the possibility that in certain regions of the
supergravity parameter space the sparticle spectrum 
can get large even for modest
values of the fine tuning parameter $\Phi_0$. This generally happens in 
regions where the loop corrections to $\mu$ are large. For example, in 
contrast to the case of small $tan\beta$ one finds that 
  for the case of large tan$\beta$ the loop corrections to $\mu$  can 
  become rather significant. 
  In this case the size of the loop corrections to $\mu$ depends sharply
  on the scale $Q_0$ where the minimization of the effective potential is 
  carried out. In fact, in this case there is generally a strong dependence
  on $Q_0$ of both the tree and the loop contributions to $\mu$ which, however,
  largely cancel in the sum, leaving the total $\mu$ with a sharply reduced
  but still non-negligible  residual $Q_0$ dependence. 
An illustration of this phenomenon is given in Fig.3. The choice of
$Q_0$ where one carries out the minimization of the effective potential 
is of importance because we can choose a value of $Q_0$ where the loop
corrections are small so that we can carry  out an analytic analysis similar 
to the one in Sec.3. (For example, for the case of Fig. 3 the loop correction
to $\mu$ is minimized at $Q_0\approx 1$ TeV). 
Generally we find the value $Q_0$ at which the loop correction to 
$\mu$ is minimized to be about the average of the
smallest and the largest sparticle masses, a value not too distant from 
 $\sqrt{m_{\tilde t_L}m_{\tilde t_R}}$,
which is typically chosen  to minimize the 2-loop correction to the 
Higgs mass\cite{carena,barger2}.  Choosing  a  value $Q_0$
 where the loop correction is 
small ($Q_0$ is typically greater than 1 TeV here), and following 
the same procedure as in Sec.3 we find that 
this time sign$(C_1(Q_0))$=-1 (see entries for the case
tan$\beta=10, 20$ in Table 1 and see also fig. 5a). There are now two distinct 
possibilities: case A
and case B which we discuss below. \\

\noindent
{\bf Case A:} This case corresponds to 
\begin{equation}
(\Phi_0+\frac{1}{4})M_Z^2-C_2'A_0^2 >0
\end{equation}
and occurs for relatively small values of $|A_0|$. Here 
 the radiative breaking equation takes the form 
\begin{equation}
\frac{m_{1/2}'^2}{\alpha^2(Q_0)}-\frac{m_0^2}{\beta^2(Q_0)}\simeq 1
\end{equation}
where 
\begin{equation}
\alpha^2=\frac{|(\Phi_0+\frac{1}{4})M_Z^2-C_2'A_0^2|}{|C_3|}
\end{equation}
and 
\begin{equation}
\beta^2=\frac{|(\Phi_0+\frac{1}{4})M_Z^2-C_2'A_0^2|}{|C_1|}
\end{equation}
The appearance of a minus sign changes intrinsically the character of the 
constraint of the electro-weak symmetry breaking. One finds now that 
unlike the previous case, where $m_0$ and $m_{1/2}$ lie on the boundary
of an ellipse for fixed $A_0$ (see Fig.4a and also Figs.1a and 2a), 
here they lie on  a hyperbola.
A diagrammatic representation of the constraint of Eq.(14)
 is given in Fig.4b-4c. The position of the apex of the hyperbola depends on
 $A_0$ as can be seen from Fig.4c. The choice of $\Phi$ itself does not 
 put an upper bound on $m_0$ and $m_{1/2}$ and consequently they  can  get 
   large for a fixed fine tuning unless other constraints 
 intervene.  Thus in this  case the rule  that the upper bounds are
 proportional to $\sqrt \Phi_0$ breaks down. In fact from Eq.(14)-(16) we 
 see that for large $m_0$ and $m_{1/2}$ one has\\
 \begin{equation}
 m_0\simeq \sqrt{\frac{|C_3|}{|C_1|}}m_{1/2}'
 \end{equation}
 and thus independent of $\Phi_0$. Thus the hyperbolae for different values
 of fine tuning  have the same asymptote independent of $\Phi_0$ as 
 illustrated in Fig.4b.\\ 
   
\noindent
{\bf Case B:} This case corresponds to  
\begin{equation}
(\Phi_0+\frac{1}{4})M_Z^2-C_2'A_0^2 <0
\end{equation}
and occurs for relatively large values of $A_0$. Here
 the radiative breaking equation takes the form 
\begin{equation}
\frac{m_0^2}{\beta^2(Q_0)}-\frac{m_{1/2}'^2}{\alpha^2(Q_0)}\simeq 1
\end{equation} 
 A diagrammatic representation of this case is given in Fig 4d. 
  As in Case A, here also $m_0$ and $m_{1/2}$ lie on 
 a hyperbola, with the position of the apex determined by 
 the value  of $A_0$. Again here as in case A the choice of 
 $\Phi_0$ itself does not control the upper bound on $m_0$ and $m_{1/2}$.
This can be seen from Fig.4d where the hyperbolae for different values of
the fine tuning have the same asymptote  independent of $\Phi_0$ just
as in case A.    	    
 	   We emphasize that the analytic analysis based  on Eqs.(14) and (19)
 	    is for 
 	   illustrative purposes
 	   only, and the results presented in this paper are obtained 
 	   including the b-quark couplings and including the full one loop
 	   corrections to $\mu$. In Fig.5b  we present a numerical analysis of 
 	   the allowed region of  $m_0$ and $m_{1/2}$. 
 	  One  finds that the cases $A_0=0$ and $A_0=500$ GeV show that
 	  $m_0$ and $m_{1/2}$ lie on  a branch 
 	  of a hyperbola and simulate the illustration of 
 	  Fig.4c.  This is what one expects for the small $A_0$ case. 
 	  Similarly for the cases $A_0=-1000$ GeV and $A_0=-2000$ GeV 
 	  in Fig.5b, $m_0$ and $m_{1/2}$ lie on a branch of a hyperbola  
 	  and simulate
 	  the illustration of the right hyperbola in 
 	   Fig.4d as is appropriate for a large negative $A_0$. 
 	   Similarly for the case $A_0=1000$ GeV in Fig.5b, $m_0$ and
 	   $m_{1/2}$ again lie on a branch of a hyperbola and simulate
 	   the illustration  of the left hyperbola in Fig.4d. 
 	   A similar analysis for $tan\beta=20$ can be found in Fig.5c. 
 	     Thus one finds that the results of the analytic 
 	    	analysis are supported by the full numerical analysis.

\section{Effects of Non-universal Soft SUSY Breaking}

The analysis of Secs. 3 and 4 
above is carried out under the assumption of universal 
soft supersymmetry boundary conditions at the GUT scale. These universal
boundary conditions arise from the assumption of  a flat Kahler 
potential. However, the framework of supergravity 
unification\cite{applied,swieca} 
allows for  
 more general Kahler structures and hence for non-universalities in 
 the soft supersymmetry breaking parameters\cite{soni,planck}. 
 In the analysis of this 
 section we shall assume universalities in the soft supersymmetry breaking 
 parameters in the first two generations 
 of matter but allow for non-universalities in the 
 Higgs sector\cite{planck,matallio,berez,nonuni} and in the
 third generation of matter\cite{planck,nonuni,dimo}. 
 It is convenient to parametrize the 
 non-universalities  in the following fashion. In the Higgs sector one has
 
 \begin{equation}
m_{H_1}^2 =  m_0^2 (1 + \delta_1),  ~~m_{H_2}^2 = m_0^2 (1 + \delta_2)
\end{equation}
Similarly in the third generation sector one has 
\begin{equation}
  m_{\tilde Q_L}^2=m_0^2(1+\delta_3), ~~m_{\tilde U_R}^2=m_0^2(1+\delta_4)
\end{equation}
A reasonable range for the non-universality parameters is $|\delta_i|\leq 1$
(i=1-4). Inclusion of non-universalities modifies the electro-weak 
symmetry breaking equation determining the parameter $\mu^2$, and leads
to corrections to the fine tuning parameter $\Phi$. One finds that with these
non-universality corrections $\Phi$ is given by 

\begin{equation}
 \Phi=-\frac{1}{4}+(\frac{m_0}{M_Z})^2 C_1'+(\frac{A_0}{M_Z})^2 C_2 +
 (\frac{m_{\frac{1}{2}}}{M_Z})^2C_3+(\frac{m_{\frac{1}{2}}
A_0}{M_Z^2})C_4+\frac{\Delta \mu^2_{loop}}{M_Z^2}
\end{equation}
where 
\begin{eqnarray}
C_1'=\frac{1}{t^2-1}(1-\frac{3 D_0-1}{2}t^2)+
\frac{1}{t^2-1}(\delta_1-\delta_2t^2-\frac{ D_0-1}{2}(\delta_2 +
\delta_3+\delta_4)t^2)\nonumber\\
+\frac{3}{5}\frac{t^2+1}{t^2-1}\frac{p S_0}{m_0^2}
\end{eqnarray}
and $C_2$, $C_3$ and $C_4$ are as defined in Eqs.(6) and (7).
Here $S_0$ is the trace anomaly term 

\begin{equation}
S_0=Tr(Ym^2)
\end{equation}
evaluated at the GUT scale $M_G$. It vanishes in the universal case since
Tr(Y)=0, but contributes when non-universalities are present. p is as
defined in Ref. \cite{nonuni}.  Numerically for $M_G=10^{16.2}$ GeV and 
 $\alpha_G=1/24$  one has $p\simeq 0.045$. Eq.(23) shows how important the
 effects of non-universalities are on $\Phi$. For a moderate value of 
 $m_0=250$ GeV the factor $(m_0/M_Z)^2$ is $\sim 7.5$ and since 
 $\delta_i\sim$O(1), $\Phi$ gets a huge shift. This means that the
 upper limits of the sparticle masses are going to be sensitively dependent
 on the magnitudes and signatures of $\delta_i$.  
  
	It is instructive to write the radiative breaking equation 
	Eq.(21) with non-universalities in a form similar to Eq.(9).
	We get 
	
	\begin{equation}
	C_1'm_0^2+C_3m'^2_{1/2}+C_2'A_0^2+\Delta \mu_{loop}^2=
	M_Z^2(\Phi+\frac{1}{4})
	\end{equation}
where $C_1'$ is defined in Eq.(23), and $C_2$ and $C_3$ are defined in
Eq.(7) and where $\Delta \mu^2_{loop}$ is the loop correction. 
  We discuss the case of non-universalities in the Higgs sector first
  and consider two extreme examples within the 
	constraint of $|\delta_i|\leq 1$(i=1-2).  These are (i) $\delta_1=1$,
	$\delta_2=-1$, and (ii)$\delta_1=-1$, $\delta_2=1$, with 
	$\delta_3=0=\delta_4$ in both cases. For case(i) we find from
	Eq.(23) that the non-universalities make a positive 
	contribution to  $C_1'$, and thus $C_1'>0$ (see Table 3).
	As for the universal case the loop corrections in this case are
	generally small.
	 Thus 
	 in this case one finds that the radiative breaking condition 
	 takes the form 
\begin{equation}
\frac{m_{1/2}'^2}{a^2}+\frac{m_0^2}{b'^2}+\frac{A_0^2}{c^2}\simeq 1
\end{equation}
where a and c are defined by Eq.(11) and $b'$ is defined by 
\begin{equation}
b'^2=M_Z^2\frac{(\Phi+\frac{1}{4})}{|C_1'|},
\end{equation}
As in the universal case (see Figs.1a, 2a and 4a) 
here also  for given fine tuning one finds that 
  $m_0$ and $m_{1/2}$ are bounded as they lie on the boundary of an ellipse. 
  	Further, $C_1'>C_1$  implies that
	a given $\Phi$ corresponds effectively to a smaller
	$\Phi_0$, and hence admits smaller values of the upper
	limits of the squark masses relative to the universal case. 
	This is what is seen in Table 4. 
	Here we find that the upper limits are generally decreased over
	the full range of $\Phi$.
	
		  	For case(ii) the situation is drastically different.
	 Here the non-universalities make a negative contribution
	  	driving $C_1'$ negative (see Table 3) and further $C_1'$
	  	remains negative in the relevant Q range (see Table 5).
Thus the radiative breaking solutions no 
  longer lie on the boundary  of an ellipse. The analysis in this case 
  is somewhat more complicated in that the loop corrections to $\mu^2$ 
  at the scale $Q=M_Z$ are large. For illustrative purposes 
  one may carry out an analysis similar to the one discussed in Sec.3 and
  go to the scale  Q=$Q_0'$, where the loop corrections to $\mu^2$ are 
  negligible. Again there are two cases and we discuss these below. \\

  \noindent
  {\bf Case C:} This case is defined by Eq.(13) and the radiative 
  symmetry breaking 
  constraint here reads
  \begin{equation}
  \frac{m'^2_{1/2}}{\alpha^2(Q_0')}-\frac{m_0^2}{\beta'^2(Q_0')}\simeq 1
  \end{equation}
  where 
  \begin{equation}
\beta'^2=\frac{|(\Phi+\frac{1}{4})M_Z^2-C_2'A_0^2|}{|C_1'|}
\end{equation}
Eq.(28) shows that the radiative symmetry breaking constraint in this 
case is a hyperbolic constraint.

    \noindent
  {\bf Case D:} This case is defined by Eq.(18) and the radiative
  symmetry  breaking 
  constraint here reads
  \begin{equation}
  \frac{m_0^2}{\beta'^2(Q_0')}- \frac{m'^2_{1/2}}{\alpha^2(Q_0')} \simeq 1
  \end{equation}
Again the radiative symmetry breaking constraint is a hyperbolic constraint.

	Cases C and D are similar to the cases A and B except that here
	$m_0$ and $m_{1/2}$ lie on a hyperbola even for small tan$\beta$
	because of the effect of the specific nature of the 
	non-universalities  in this case. Thus here it is the 
	non-universalities which  transform the radiative breaking 
	equation from an ellipse to a hyperbola. Of course $m_0$ and
	$m_{1/2}$ do not become arbitrarily large, since
	eventually other constraints set in and 
limit the allowed values of $m_0$ and $m_{1/2}$. Results of the analysis 
are given in  Fig.(6). One finds that $m_0$ and $m_{1/2}$
 indeed can become large for a fixed fine tuning.

	 To understand the effects of the non-universalities in the third
	 generation in comparison to the non-universalities  in the Higgs 
	 sector it is useful to express $\Delta \Phi$ in the following 
	 alternate form 

\begin{equation}
\Delta\Phi=
\frac{1}{t^2-1}(\delta_1-(1-\frac{1}{2} (\frac{m_t}{m_f})^2)\delta_2t^2
+\frac{1}{2} (\frac{m_t}{m_f})^2(
\delta_3+\delta_4)t^2) (\frac{m_0}{M_Z})^2
+\frac{3}{5}\frac{t^2+1}{t^2-1}\frac{p S_0}{M_Z^2}
\end{equation} 
Since $m_t<m_f$ one has
$(1-\frac{1}{2} (\frac{m_t}{m_f})^2)>0$ which implies that 
the effect of a negative(positive) $\delta_2$ 
can be
simulated by a positive(negative) value of $\delta_3$ or
 by a positive(negative) value of $\delta_4$.  
 This correlation can be seen to hold by a comparison of Tables 4 and 6.
As in the case of Table 4 where  a positive $\delta_1$ and a negative
 $\delta_2$ leads to  lowering
of the upper limits on squark  masses, we find that a positive 
$\delta_3$ or  a positive $\delta_4$  produces  a similar effect.
 The analysis of Table 6 where  we choose ($\delta_1,\delta_2,\delta_3,
 \delta_4$)=(0,0,1,0) supports this observation. 
 A similar correlation can be made between the case of $\delta_1<0,
 \delta_2>0$ and the case $\delta_3+\delta_4<0$ by the comparison 
 given above.    We note, 
however, that the effects of non-universalities in the Higgs sector and in the 
third generation sector are not identical in every respect as they 
enter in different ways in other parts of the spectrum. However, 
the gross features of the upper limits of squarks in Table 6  can be 
understood by the rough comparison given above.

	A comparison of Tables 2,4, and 6
	 shows that the non-universalities have
	a remarkable effect on the upper limits of sparticle masses. One 
	finds that the upper limits on the sparticle masses  can 
	increase or decrease dramatically depending on the type of
	non-universality included in the analysis. 
	The prospects for the observation of 
	sparticles at colliders are thus significantly affected. 
	For the case of
	Table 4 and 6 one finds that the sparticle spectrum falls below 
	1 TeV in the  range  $tan\beta\leq 5, \Phi\leq 20$. Thus 
	in this case the gluino and  the 
	squarks should be discovered at the LHC and all of  the other
	sparticles should also be discovered over most of the mass ranges
	in Table 4.
	In contrast for the case of non-universality of Table.6 
	 we find that the nature of 
	non-universal
	contribution is such that squark masses can exceed 
	the discovery potential of even the LHC.
	The analysis given above is for $\mu<0$. A similar analysis holds
	with essentially the same general conclusions for the $\mu>0$ 
	case.

\section{Upper Limit on the Higgs Mass}
 	
 	One of the most interesting part of our analysis concerns the
 	dependence of the 
 	Higgs mass upper limits on $\Phi_0$. For the analysis of the 
 	Higgs mass upper limits we have taken account of the one loop 
 	corrections to the masses and further chosen the scale Q which 
 	minimizes the two loop corrections\cite{carena,barger2}.
 	 For tan$\beta=2$ the upper limit on the Higgs mass
	increases from 60 GeV at $\Phi_0$=2.5 GeV to 86 GeV at 
	$\Phi_0$=20. Further from the successive entries in this case
	in Table 2 we observe that  in each of the cases where an 
	increment in the Higgs mass occurs, one requires a significant
	increase in the value of $\Phi_0$. 
	 The same general pattern is repeated for larger 
 	values of tan$\beta$. Thus for tan$\beta=5$ the
 	Higgs mass  increases from  97 GeV to 116 GeV as $\Phi_0$
 	increases from 2.5 to 20. 
	In Fig.7 we exhibit the upper bound on the Higgs  mass as a 
	function of tan$\beta$. From the analysis of Table 2 and Fig.7 one
	 can draw 
	the general conclusion that  the Higgs mass upper limit is  a 
	sensitive function of tan$\beta$ and  $\Phi_0$.
	For values  of tan$\beta$ near the low end, i.e. tan$\beta\approx 2$,
	the upper limit of the Higgs mass lies below 85-90 GeV for any 
	reasonable range of fine tuning, i.e. $\Phi_0\leq20$. 
	This is a rather strong result. Thus if the low tan$\beta$ region
	of $b-\tau$ unification turns out to  be the correct scenario
	then our analysis implies the existence of a Higgs mass
	below 85-90 GeV for any reasonable range of fine tuning. 
	This scenario will be completely tested 
	at LEPII which can allow  coverage of the Higgs mass  up to 
	$m_h\approx 95$ GeV with $\sqrt s= 192$ GeV. 
	If no Higgs  is seen at LEPII then a high degree of fine tuning,
	i.e. $\Phi_0>20$, is indicated on the low tan$\beta$ end of 
	$b-\tau$ unification.
		
	  Further, the analysis also indicates that in order to 
	  approach the maximum allowed Higgs mass one needs to have 
	  a high degree of fine tuning. 
	  In particular from Table 2 and Fig.7  we see that going beyond 
	  120 GeV in the 
  	Higgs mass requires a value of $\Phi_0$ on the high side,
  	preferably 10 and 20. 
	The strong correlation of the Higgs mass upper limits
 	with the value of $\Phi_0$ has important implications for 
 	sparticle masses. 
	 Thus if the Higgs mass turns out
 		 to lie  close to its allowed upper limit then
 		 a larger value of $\Phi_0$ would be indicated. In turn
 		 a large $\Phi_0$ would point to a heavy 
 		 sparticle spectrum. At TeV33 with 25fb$^{-1}$ of 
 		 integrated luminosity Higgs mass up to 120 GeV 
 		 will be probed.  A non-observation of the light Higgs 
 		 in this mass range will imply that one needs a high degree
 		 of fine tuning which would point in the direction
 		 of  heavy sparticle masses.
 	These  results are in general agreement with the  analysis 
 	of Ref.\cite{ACR} which arrived at much the same conclusion using a 
 	very different criterion of fine tuning. In particular the 
 	analysis of Ref.\cite{ACR} also found that the non-observation
 	of the Higgs mass below 120 GeV will imply a heavy spectrum.

\section{Fine-tuning limits from the current experimental data}
One may put limits on the fine tuning parameter using the current
experimental data on sparticle searches at colliders\cite{lep,d0}. 
The result
of this analysis is presented in Table 7. For low tan$\beta$ the strongest 
lower limits on the fine tuning parameter arise from the lower 
limits on the Higgs mass. In Table 7 we have used the experimental
lower limits on the Higgs mass from the four detectors at LEP, i.e.,
the L3, OPAL, ALEPH, and DELPHI\cite{lep}, to obtain lower limits on $\Phi$ 
for values of tan$\beta$ from 2 to 20. As expected one finds that
the strongest limit on $\Phi$ arises for the smallest tan$\beta$,
and the constraint on $\Phi$ falls rapidly for larger tan$\beta$.
Thus for tan$\beta$ greater than 5 the lower limit on $\Phi$ 
already drops below 2 which is not a stringent fine tuning constraint.
Lower limits on $\Phi$ from the current data on the lower limits 
on the neutralino, the chargino, the stop, the heavy squarks, and
the gluino are also analysed in Table 7. One finds that here the 
current lower limits on the chargino mass produce the stongest
lower limit on $\Phi$. For tan$\beta$ of 2, the lower limit on 
$\Phi$ from the Higgs sector is still more stringent constraint
than the lower limit constraint from the chargino sector. However,
for tan$\beta$=5 the constraint from the chargino sector becomes 
more stringent than the constraint from the Higgs sector. These
constraints on the fine tuning will become even more stringent 
after LEP II completes its runs and if supersymmetric 
particles do not become visible.

\section{Conclusions}
In this paper we have analyzed the naturalness bounds on sparticle masses
within the framework of radiative breaking of the electro-weak symmetry
for minimal supergravity models and for non-minimal models with 
non-universal soft SUSY breaking terms. For the case of minimal 
supergravity it is found that for small values of tan$\beta$, i.e., 
tan$\beta\leq 5$
 and a reasonable range of fine tuning, i.e., $\Phi\leq 20$, the 
 allowed values of $m_0$ and $m_{1/2}$ lie on the surface of an ellipsoid
 with the  radii  determined by the value of fine tuning. Specifically for 
 the case tan$\beta=2$ it is found that the upper limits on the
 gluino and squark masses in minimal supergravity lie within 1 TeV and the 
 light Higgs mass lies below 90 GeV
for $\Phi_0\leq 20$. For
tan$\beta\leq 5$ the upper limits of the sparticle masses all still  
lie within the reach  of the LHC for the same range of $\Phi_0$. 
The analysis shows that the upper limits of sparticle masses are very
sensitive functions of tan$\beta$. As values of tan$\beta$ become large
the loop corrections to $\mu$  become large and the nature of the 
radiative breaking equation can change, i.e., $m_0$ and $m_{1/2}$ may
not lie on the surface of an ellipsoid. Thus it is found that there
exist regions of the parameter space for large tan$\beta$ where the upper bounds
on the sparticle masses can get very large even for reasonable values of
fine tuning. 

 We have also analyzed the effects of non-universalities in the Higgs sector
 and in the third generation sector on the upper limits on the 
 sparticle masses.  It  is found that 
 non-universalities have a very significant effect on the overall size of
 the  sparticle mass upper limits.
 Thus we find that the case (i) $\delta_1>0$ or $\delta_2<0$ and 
 $\delta_3=0=
 \delta_4$  has the effect of decreasing the upper limits
 on the squark masses, and in contrast the case (ii) 
 $\delta_1<0$ or $\delta_2>0$ and $\delta_3=0=
 \delta_4$  has the effect of increasing the upper limits
 on the squark masses. Remarkably for  $\delta_1=1$, $\delta_2=-1$ and
  $\delta_3=0=\delta_4$  
	all of the sparticle masses lie below  1 TeV 
	for tan$\beta\leq 5$ and $\Phi\leq 20$ 
	because of the 
	non-universality effects. In this case the sparticles would
	not escape detection at the LHC. However, for the
	case  $\delta_1=-1$, $\delta_2=1$ and
  $\delta_3=0=\delta_4$  there is an opposite effect
   and the non-universalities raise the 
	upper limits of the sparticle masses. Here 
	for the same range  of tan$\beta$, i.e.,
	 tan$\beta\leq 5$   
	the first and second generation squark masses can reach 
	approximately 3 TeV for $\Phi\leq 10$ (4-5 TeV for 
	$\Phi\leq 20$)  
	and consequently these sparticles may escape
	detection even at the LHC. Similar effects occur for the 
	non-universalities in the third generation sector. 
	Thus non-universalities
	have  important implications for the detection of supersymmetry
	at colliders.

  Finally, it is  found that the upper limit
on the Higgs mass is a very sensitive function of tan$\beta$ in the 
region of low $tan\beta$ and moving the upper limit beyond
120 GeV towards its maximally allowed value will require a high degree of
fine tuning.  In turn large fine tuning would result in a 
corresponding  upward movement of the upper limits of  other sparticle masses.
 Thus a non-observation of the Higgs at the upgraded Tevatron with an 
 integrated luminosity of 25$fb^{-1}$, would imply a high degree of 
 fine tuning and point to the possibility of a heavy sparticle spectrum.

{\bf Acknowledgements}\\
Fruitful discussions with  Richard Arnowitt, Howard Baer and Haim 
Goldberg are acknowledged. 
This research was supported in part by NSF grant number PHY-96020274.


\begin{center}						
\begin{table}

\begin{center}						
\begin{tabular}{|c||c|c|c|c|c|}
\hline
\hline
\multicolumn{6}{|c|}{Scale dependence of $C_1$ -- $C_4$ }	\\
\hline
$\tan{\beta}$ 	& $Q(GeV)$	& $C_1$	& $C_2$	& $C_3$	& $C_4$	\\
\hline
2       &      91.2    &    0.7571   &    0.0711   &     4.284   &    0.3119  \\
		& 2000	& 0.6874	& 0.0879	& 2.851	& 0.3073	\\
		& 4000	& 0.6702	& 0.0918	& 2.607	& 0.3055	\\
		& 6000	& 0.6598	& 0.0941	& 2.474	& 0.3043	\\
		& 8000	& 0.6523	& 0.0957	& 2.384	& 0.3034	\\
		& 10000	& 0.6464	& 0.0970	& 2.316	& 0.3026	\\
\hline
5      &      91.2  &    0.14212   &    0.1024   &     2.871   &    0.4491  \\
       &     500   &    0.09016   &    0.1099   &     2.200   &    0.4245  \\
       &    1000   &    0.06843   &    0.1126   &     1.973   &    0.4138  \\
       &    1500   &    0.05558   &    0.1142   &     1.851   &    0.4074  \\
       &    2000   &    0.04639   &    0.1152   &     1.768   &    0.4028  \\
       &    2500   &    0.03924   &    0.1160   &     1.706   &    0.3992  \\
       &    3000   &    0.03336   &    0.1166   &     1.657   &    0.3962  \\
       &    3500   &    0.02838   &    0.1172   &     1.617   &    0.3937  \\
       &    4000   &    0.02406   &    0.1176   &     1.583   &    0.3914  \\
       &    4500   &    0.02023   &    0.1180   &     1.553   &    0.3895  \\
       &    5000   &    0.01680   &    0.1184   &     1.527   &    0.3877  \\
\hline
10       &      91.2   &    0.0756   &    0.1040   &     2.710   &    0.4561  \\
       &     250   &    0.0446   &    0.1081   &     2.305   &    0.4397  \\
       &     500   &    0.0230   &    0.1108   &     2.062   &    0.4280  \\
       &     750   &    0.0102   &    0.1122   &     1.931   &    0.4211  \\
       &    1000   &    0.0011   &    0.1132   &     1.843   &    0.4160  \\
       &    1250   &   -0.0060   &    0.1140   &     1.778   &    0.4121  \\
       &    1500   &   -0.0118   &    0.1146   &     1.726   &    0.4089  \\
       &    1750   &   -0.0167   &    0.1151   &     1.683   &    0.4061  \\
       &    2000   &   -0.0210   &    0.1155   &     1.646   &    0.4037  \\
       &    2500   &   -0.0281   &    0.1162   &     1.587   &    0.3997  \\
       &    3000   &   -0.0341   &    0.1167   &     1.540   &    0.3964  \\
\hline	
20     &     250   &    0.02850   &    0.1084   &     2.269   &    0.4406  \\
       &     500   &    0.00685   &    0.1109   &     2.029   &    0.4286  \\
       &     750   &   -0.00592   &    0.1123   &     1.899   &    0.4214  \\
       &    1000   &   -0.01504   &    0.1133   &     1.812   &    0.4162  \\
       &    1250   &   -0.02213   &    0.1140   &     1.747   &    0.4122  \\
       &    1500   &   -0.02795   &    0.1146   &     1.695   &    0.4089  \\
       &    1750   &   -0.03288   &    0.1150   &     1.653   &    0.4061  \\
       &    2000   &   -0.03716   &    0.1154   &     1.617   &    0.4036  \\
       &    2500   &   -0.04433   &    0.1161   &     1.558   &    0.3995  \\
       &    3000   &   -0.05020   &    0.1166   &     1.511   &    0.3961  \\
\hline
\hline
\end{tabular}
\end{center}

\small{\caption{
	The  scale dependence of $C_1(Q)$ - $C_4(Q)$
	 for minimal 
	supergravity when  
	$m_t=175~GeV$
	 for  $tan{\beta}=$ 2, 5, 10 and 20.
	}}
\end{table}
\end{center}

\begin{center}						
\begin{table}

\begin{center}						
\begin{tabular}{|c|c||c|c|c|c|c|c|c|c|c|c|c|c|}
\hline
\hline
\multicolumn{14}{|c|}{Minimal Supergravity 	\qquad $\mu <0$}	\\
\hline
$\tan{\beta}$ & $\Phi$	& H	& $\tilde u_l$	& $\tilde e_l$	& $\tilde e_r$	& $\tilde t_1$	& $\tilde t_2$	& $\tilde g$	& h	
&  $\tilde \chi^\pm_1$ & $\tilde \chi^\pm_2$ &   $\tilde \chi^0_1$ &	$m_0$\\
\hline  
2	& 5	& 326	& 290	& 212	& 207	& 264	& 325	& 316	& 69	& 102	& 224	& 48	& 204	\\
	& 10	& 479	& 419	& 320	& 315	& 353	& 429	& 459	& 78	& 139	& 303	& 67	& 315	\\
	& 20	& 687	& 598	& 463	& 459	& 483	& 579	& 649	& 86	& 190	& 419	& 94	& 459	\\
\hline  
5	& 2.5	& 318	& 352	& 292	& 285	& 265	& 352	& 295	& 97	& 77	& 180	& 42	& 282	\\
	& 5	& 594	& 589	& 560	& 556	& 365	& 507	& 425	& 103	& 114	& 232	& 60	& 556	\\
	& 10	& 930	& 906	& 888	& 884	& 510	& 744	& 610	& 109	& 167	& 309	& 86	& 886	\\
	& 20	& 1417	& 1381	& 1368	& 1365	& 742	& 1113	& 873	& 116	& 243	& 423	& 123	& 1368	\\
\hline	
10	& 2.5	&416	&464	&403	&393	&323	&431	&316	&106	&73	&184	&42	&395	\\
	& 5	&3702	&4089	&3914	&3887	&2311	&3311	&1272	&136	&190	&382	&158	&3920	\\
	& 10	&5963	&6714	&6365	&6318	&3855	&5428	&2776	&144	&283	&797	&273	&6370	\\
	& 20	&8875	&10536	&9622	&9527	&6170	&8616	&4945	&150	&404	&1409	&400	&9596	\\
\hline	
20	& 2.5	&1889	&2136	&2044	&2003	&1202	&1697	&566	&128	&104	&214	&69	&2080	\\
	& 5	&3581	&4198	&3906	&3827	&2480	&3383	&1764	&138	&194	&515	&178	&3980	\\
	& 10	&5540	&6585	&6114	&5978	&3893	&5270	&3124	&145	&282	&895	&274	&6210	\\
	& 20	&8007	&10092	&8954	&8734	&6078	&8167	&5322	&151	&403	&1516	&399	&9060	\\
\hline
\hline
\end{tabular}
\end{center}

\small{\caption{
	The  upper bound on sparticle masses for minimal 
	supergravity when  
	$m_t=175 ~GeV$ and $\mu <0$ for different values of tan$\beta$ and
	fine tuning measure $\Phi_0$. All the masses are in GeV.
	}}
\end{table}
\end{center}

\begin{center}						
\begin{table}
\begin{center}						
\begin{tabular}{|c|c|c|c|}
\hline
\hline
$\tan{\beta}$	&$\delta_1$   & $\delta_2$ 	& $C'_1$	\\
\hline
2   &  -1.0   &    1.0   &    -0.341  \\
   &  -0.75   &    0.75   &    -0.067  \\
   &  -0.5   &    0.5   &     0.208  \\
   &  -0.25   &    0.25   &     0.483  \\
   &   0.0   &    0.0   &     0.757  \\
   &   0.25   &   -0.25   &     1.032  \\
   &   0.5   &   -0.5   &     1.306  \\
   &   0.75   &   -0.75   &     1.581  \\
   &   1.0   &   -1.0   &     1.855  \\
\hline
5   &  -1.0   &    1.0   &    -0.572  \\
   &  -0.75   &    0.75   &    -0.393  \\
   &  -0.5   &    0.5   &    -0.215  \\
   &  -0.25   &    0.25   &    -0.036  \\
   &   0.0   &    0.0   &     0.142  \\
   &   0.25   &   -0.25   &     0.321  \\
   &   0.5   &   -0.50   &     0.499  \\
   &   0.75   &   -0.75   &     0.677  \\
   &   1.0   &   -1.0   &     0.856  \\
\hline
10   &  -1.0   &    1.0   &    -0.597  \\
   &  -0.75   &    0.75   &    -0.429  \\
   &  -0.5   &    0.5   &    -0.261  \\
   &  -0.25   &    0.25   &    -0.092  \\
   &   0.0   &    0.0   &     0.076  \\
   &   0.25   &   -0.25   &     0.244  \\
   &   0.5   &   -0.5   &     0.412  \\
   &   0.75   &   -0.75   &     0.580  \\
   &   1.0   &   -1.0   &     0.748  \\
\hline
20   &  -1.0   &    1.0   &    -0.603  \\
   &  -0.75   &    0.75   &    -0.437  \\
   &  -0.5   &    0.5   &    -0.272  \\
   &  -0.25   &    0.25   &    -0.106  \\
   &   0.0   &    0.0   &     0.060  \\
   &   0.25   &   -0.25   &     0.225  \\
   &   0.5   &   -0.5   &     0.391  \\
   &   0.75   &   -0.75   &     0.556  \\
   &   1.0   &   -1.0   &     0.722  \\
\hline
\hline
\end{tabular}
\end{center}
\small{\caption{$C_1'(M_Z)$ for different values   
	of $\delta_1$ and $\delta_2$ when   
	$m_t=175 ~GeV$ for  $tan{\beta}=$ 2, 5, 10 and 20.
	}}
\end{table}
\end{center}

\begin{center}
\begin{table}

\begin{center}
\begin{tabular}{|c|c||c|c|c|c|c|c|c|c|c|c|c|c|}
\hline
\hline
\multicolumn{14}{|c|}{Non-universal case:~~$(\delta_1,\delta_2,
\delta_3,\delta_4)=(1,-1,0,0),	\qquad \mu <0$}	\\
\hline
$\tan{\beta}$ & $\Phi$	& H	& $\tilde u_l$	& $\tilde e_l$	& $\tilde e_r$	& $\tilde t_1$	& $\tilde t_2$	& $\tilde g$	& h	
&  $\tilde \chi^\pm_1$ & $\tilde \chi^\pm_2$ &   $\tilde \chi^0_1$ &	$m_0$\\
\hline  
2	& 5	& 313	& 291	& 148	& 140	& 267	& 328	& 319	& 70	& 104	& 225	& 49	& 134	\\
	& 10	& 457	& 419	& 220	& 207	& 354	& 430	& 459	& 79	& 140	& 304	& 68	& 205	\\
	& 20	& 655	& 601	& 317	& 296	& 488	& 585	& 656	& 87	& 193	& 420	& 95	& 299	\\
\hline  
5	& 2.5	& 213	& 274	& 133	& 121	& 243	& 336	& 301	& 95	& 79	& 181	& 44	& 109	\\
	& 5	& 356	& 391	& 221	& 228	& 324	& 430	& 429	& 103	& 115	& 233	& 62	& 204	\\
	& 10	& 535	& 569	& 334	& 312	& 462	& 580	& 620	& 110	& 170	& 310	& 89	& 315	\\
	& 20	& 775	& 841	& 485	& 451	& 677	& 820	& 915	& 116	& 254	& 425	& 130	& 462	\\
\hline  
10	& 2.5	& 203	& 285	& 129	& 103	& 246	& 349	& 316	& 104	& 74	& 185	& 43	& 93	\\
	& 5	& 371	& 406	& 234	& 216	& 333	& 447	& 446	& 110	& 112	& 237	& 62	& 215	\\
	& 10	& 559	& 583	& 357	& 332	& 471	& 598	& 637	& 116	& 169	& 314	& 90	& 338	\\
	& 20	& 810	& 843	& 520	& 483	& 680	& 828	& 920	& 122	& 251	& 427	& 130	& 496	\\
\hline  
20	& 2.5	& 216	& 286	& 125	& 69	& 242	& 350	& 315	& 105	& 69	& 186	& 40	& 75	\\
	& 5	& 383	& 409	& 236	& 203	& 330	& 451	& 448	& 111	& 109	& 237	& 60	& 216	\\
	& 10	& 577	& 590	& 357	& 323	& 472	& 604	& 646	& 117	& 167	& 315	& 89	& 343	\\
	& 20	& 843	& 854	& 519	& 479	& 687	& 835	& 932	& 122	& 252	& 428	& 130	& 508	\\
\hline
\hline
\end{tabular}
\end{center}

\small{\caption{
        The  upper bounds on sparticle masses for the case of 
	non-universalities in the Higgs sector when 
	 $(\delta_1,\delta_2)=(1,-1)$, $\delta_3=0=\delta_4$,
	$m_t=175 GeV$, and  $\mu <0$ for different values of $tan\beta$
	and $\Phi$. All the masses are in GeV.
	}}
\end{table}
\end{center}

\begin{center}						
\begin{table}
\begin{center}						
\begin{tabular}{|c|c|c|c|c|c|}
\hline
\hline
\multicolumn{6}{|c|}{ $(\delta_1,\delta_2,\delta_3,\delta_4)=(-1, 1, 0, 0)$} \\
\hline
$\tan{\beta}$	&$Q(GeV)$	& $C'_1$	& $C_2$	& $C_3$	& $C_4$	\\
\hline
2      &    91.2   &    -0.341   &     0.071   &     4.284   &     0.312  \\
       &     250   &     -0.363   &    0.0765   &     3.742   &    0.3110  \\
       &     500   &     -0.378   &    0.0802   &     3.415   &    0.3101  \\
       &     750   &     -0.387   &    0.0825   &     3.239   &    0.3094  \\
       &    1000   &     -0.394   &    0.0841   &     3.119   &    0.3089  \\
       &    1250   &     -0.399   &    0.0853   &     3.030   &    0.3084  \\
       &    1500   &     -0.404   &    0.0863   &     2.959   &    0.3080  \\
       &    1750   &     -0.408   &    0.0872   &     2.901   &    0.3077  \\
       &    2000   &     -0.411   &    0.0879   &     2.851   &    0.3073  \\
\hline	
5      &    91.2   &     -0.572   &    0.1024   &     2.871   &    0.4491  \\
       &     250   &     -0.602   &    0.1069   &     2.452   &    0.4348  \\
       &     500   &     -0.624   &    0.1099   &     2.200   &    0.4245  \\
       &     750   &     -0.636   &    0.1115   &     2.064   &    0.4183  \\
       &    1000   &     -0.645   &    0.1126   &     1.973   &    0.4138  \\
       &    1250   &     -0.652   &    0.1135   &     1.905   &    0.4103  \\
       &    1500   &     -0.658   &    0.1142   &     1.851   &    0.4074  \\
       &    1750   &     -0.663   &    0.1147   &     1.806   &    0.4049  \\
       &    2000   &     -0.667   &    0.1152   &     1.768   &    0.4028  \\
\hline	
10     &    91.2   &     -0.597   &    0.1040   &     2.710   &    0.4561  \\
       &     250   &     -0.628   &    0.1081   &     2.305   &    0.4397  \\
       &     500   &     -0.649   &    0.1108   &     2.062   &    0.4280  \\
       &     750   &     -0.662   &    0.1122   &     1.931   &    0.4211  \\
       &    1000   &     -0.671   &    0.1132   &     1.843   &    0.4160  \\
       &    1250   &     -0.678   &    0.1140   &     1.778   &    0.4121  \\
       &    1500   &     -0.684   &    0.1146   &     1.726   &    0.4089  \\
       &    1750   &     -0.689   &    0.1151   &     1.683   &    0.4061  \\
       &    2000   &     -0.693   &    0.1155   &     1.646   &    0.4037  \\
\hline	
20     &    91.2   &     -0.603   &    0.1043   &     2.671   &    0.4575  \\
       &     250   &     -0.634   &    0.1084   &     2.269   &    0.4406  \\
       &     500   &     -0.655   &    0.1109   &     2.029   &    0.4286  \\
       &     750   &     -0.668   &    0.1123   &     1.899   &    0.4214  \\
       &    1000   &     -0.677   &    0.1133   &     1.812   &    0.4162  \\
       &    1250   &     -0.684   &    0.1140   &     1.747   &    0.4122  \\
       &    1500   &     -0.690   &    0.1146   &     1.695   &    0.4089  \\
       &    1750   &     -0.695   &    0.1150   &     1.653   &    0.4061  \\
       &    2000   &     -0.699   &    0.1154   &     1.617   &    0.4036  \\
\hline
\hline
\end{tabular}
\end{center}
\small{\caption{The  scale dependence of $C_1'(Q)$,$C_2(Q)$ -- $C_4(Q)$ 
	for 
	$(\delta_1,\delta_2,\delta_3,\delta_4)=(-1, 1, 0, 0)$ 
	for $m_t=175 ~GeV$ and $tan{\beta}=$ 2, 5, 10 and 20.
	}}
\end{table}
\end{center}

\begin{center}
\begin{table}

\begin{center}
\begin{tabular}{|c|c||c|c|c|c|c|c|c|c|c|c|c|c|}
\hline
\hline
\multicolumn{14}{|c|}{$(\delta_1,\delta_2,\delta_3,\delta_4)=(0,0,1,0),	\qquad \mu <0$}	\\
\hline
$\tan{\beta}$ & $\Phi$	& H	& $\tilde u_l$	& $\tilde e_l$	& $\tilde e_r$	& $\tilde t_1$	& $\tilde t_2$	& $\tilde g$	& h	
&  $\tilde \chi^\pm_1$ & $\tilde \chi^\pm_2$ &   $\tilde \chi^0_1$ &	$m_0$\\
\hline  
2	& 5	& 290	& 292	& 162	& 140	& 266	& 326	& 319	& 70	& 104	& 225	& 49	& 147	\\
	& 10	& 418	& 419	& 242	& 211	& 351	& 429	& 456	& 79	& 139	& 304	& 68	& 226	\\
	& 20	& 597	& 601	& 349	& 306	& 486	& 583	& 656	& 87	& 193	& 420	& 95	& 331	\\
\hline  
5	& 2.5	& 198	& 275	& 151	& 125	& 244	& 337	& 301	& 95	& 79	& 181	& 44	& 126	\\
	& 5	& 325	& 391	& 258	& 223	& 325	& 430	& 429	& 103	& 115	& 233	& 62	& 239	\\
	& 10	& 485	& 559	& 389	& 340	& 453	& 573	& 611	& 110	& 168	& 310	& 87	& 368	\\
	& 20	& 702	& 807	& 571	& 501	& 652	& 791	& 880	& 116	& 245	& 424	& 125	& 546	\\
\hline  
10	& 2.5	& 194	& 291	& 144	& 110	& 247	& 349	& 316	& 104	& 74	& 185	& 43	& 110	\\
	& 5	& 342	& 422	& 279	& 239	& 333	& 445	& 446	& 110	& 112	& 237	& 62	& 259	\\
	& 10	& 514	& 606	& 428	& 371	& 471	& 599	& 637	& 116	& 169	& 314	& 90	& 407	\\
	& 20	& 747	& 870	& 629	& 549	& 680	& 828	& 920	& 122	& 251	& 427	& 130	& 604	\\
\hline  
20	& 2.5	& 213	& 293	& 134	& 77	& 245	& 352	& 316	& 106	& 70	& 187	& 41	& 88	\\
	& 5	& 360	& 431	& 281	& 229	& 331	& 452	& 449	& 112	& 110	& 238	& 61	& 261	\\
	& 10	& 541	& 621	& 433	& 364	& 473 	& 605	& 646	& 118	& 168	& 316	& 90	& 416	\\
	& 20	& 804	& 894	& 638	& 543	& 688	& 836	& 932	& 123	& 252	& 429	& 131	& 620	\\
\hline
\hline
\end{tabular}
\end{center}

\small{\caption{
	The  upper bound on sparticle masses for 
	non-universalities in the third generation when 
	$(\delta_3,\delta_4)=(1,0)$, 
	$m_t=175 GeV$, and $\mu <0$ for different values of tan$\beta$
	and $\Phi$. All the masses are in GeV.
	}}
\end{table}
\end{center}


\begin{center}				
\begin{table}

\begin{center}	
\begin{tabular}{|c|c|c|}
\hline
\hline
\multicolumn{3}{|c|}{$\sqrt{s}=183$ GeV LEP 95\% C.L.lower bound on $m_{h}$ }	\\
\hline
$\tan{\beta}$ & mass lower bdd (GeV)&$\Phi$($\mu<0$)	\\
\hline
2	&86 (L3)		&	20	\\
	&74 (OPAL scan B)	&	8	\\
	&88 (ALEPH)		&	23	\\
	&84 (DELPHI)		&	18	\\
\hline
5	&72 (L3)	&	$<2$	\\
	&71 (OPAL scan B)&	\\
	&73 (ALEPH)	&	\\
	&76 (DELPHI)	&	\\
\hline
10	&72 (L3)	&	$<2$	\\
	&70 (OPAL scan B)&		\\
	&76 (ALEPH)&	\\
	&75 (DELPHI)&	\\
\hline
20	&71 (L3)	&	$<2$	\\
	&70 (OPAL scan B)&	\\
	&76 (ALEPH)&	\\
	&76 (DELPHI)&	\\
\hline
\hline
\multicolumn{3}{|c|}{$\sqrt{s}=183$ GeV LEP 95\% C.L. lower bounds  on various sprticles masses }	\\
\hline
 Particle &   mass lower bdd (GeV)&$\Phi$($\mu<0$)	\\
\hline
		&	24   independent of $m_0$ (DELPHI)	& 	\\
$\chi^0$	&	14   any $m_0$ (ALEPH) 			&$<1.5$ for $\tan{\beta} \geq 2$		\\
		&	27 for $\tan{\beta}=2$ (L3)		& $<1.5$	\\
\hline
$\chi^\pm$	&	51 (ALEPH) 	& $<1.5$ for $\tan{\beta} \geq 2$	\\
\hline
$\tilde t$  &  $\tilde t \rightarrow c  \chi$	   $m_{\tilde t}>74$ (ALEPH) &$<1.5$ for $\tan{\beta} \geq 2$	\\
	&   $\tilde t \rightarrow b l \nu  \chi$   $m_{\tilde t}>82$ (ALEPH) &	\\
\hline
\hline
\multicolumn{3}{|c|}{ 95\% C.L. lower bounds on various sprticles masses from 
ref\cite{d0} }	\\
\hline
 Particle &   mass lower bdd (GeV)&$\Phi$($\mu<0$)	\\
\hline
$\chi^\pm$	& $m_{\chi^\pm}>45$,	.66 pb		&	$< 1.5$ \\
		&		&	\\
	&$m_{\chi^\pm} >124$,	.01 pb	& $\Phi >8$, $\tan{\beta}=2$	\\
	&		&	 $\Phi > 5.8$, $\tan{\beta}=5$	\\
\hline
$\tilde q$  $\tilde g$	& $m_{\tilde g}>230$, heavy squarks	&$\Phi>2.7$, $\tan{\beta}=2$ \\
		&		&	\\
		&$m_{\tilde q ,\tilde g}>260$, $m_{\tilde q}=m_{\tilde g}$	&$\Phi>4.0$, $\tan{\beta}=2$ \\
		&		&	 $\Phi<1.8$, $\tan{\beta} \geq 5$ \\
		&		&	\\
		&$m_{\tilde q}>219$, heavy gluinos		& $\Phi>2.8$,	$\tan{\beta}=2$ \\
		&		&$\Phi<1.8$,	$\tan{\beta} \geq5$  \\
\hline
\hline
\end{tabular}
\end{center}

\small{\caption{
Current experimental lower bounds on masses of the lightest Higgs and various 
sparticles from  LEP and the Tevatron.  Corresponding fine-tunings ($\mu<0$) are also shown.
	}}
\end{table}
\end{center}

\newpage
\begin{figure}[htbp]
\begin{minipage}[t]{3.0in}
	\includegraphics[angle=270,width=3.1in]{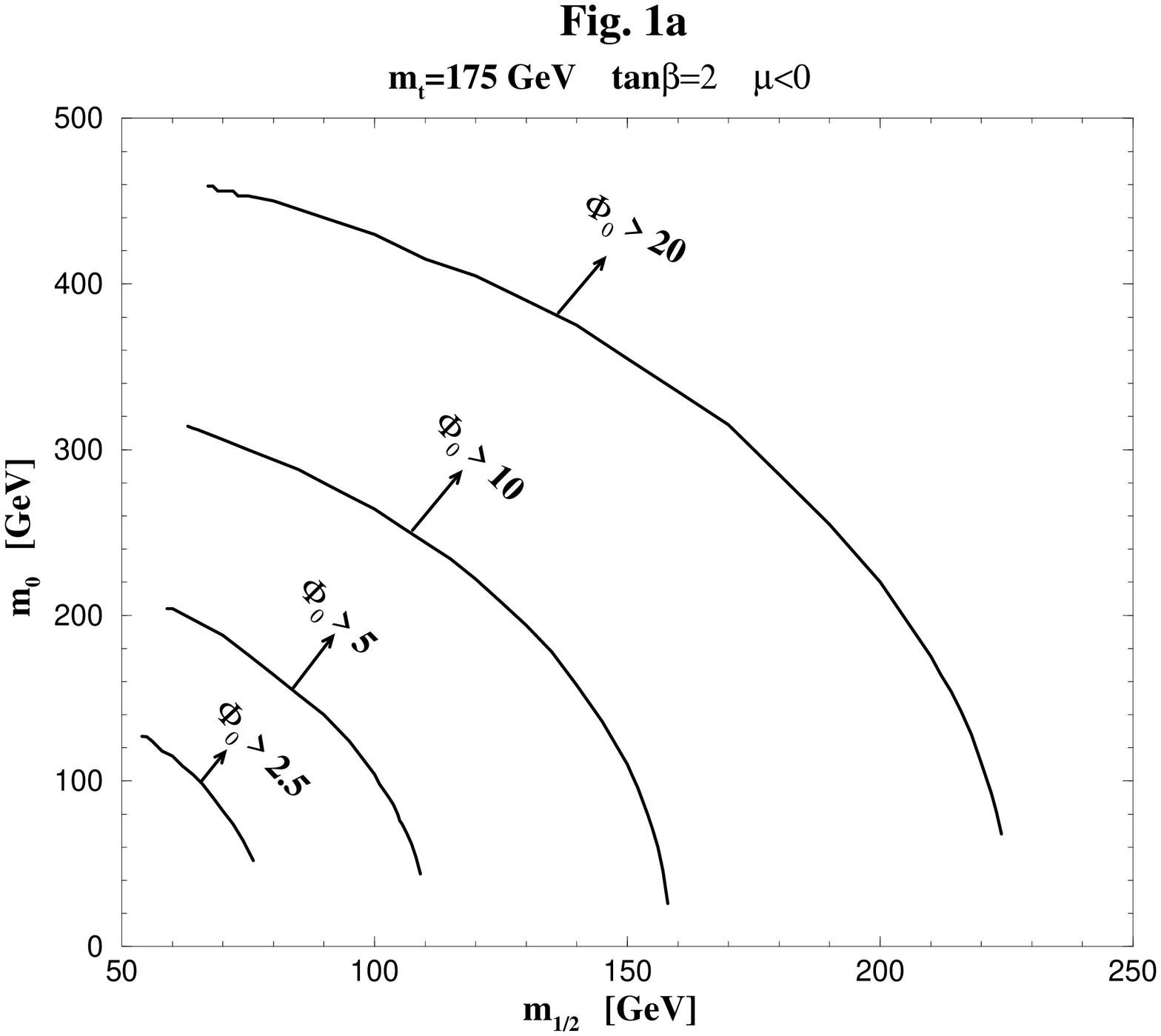} 
	\small{
	Fig.1a. Contour plot of the upper limit in the $m_0 - m_\frac{1}{2}$  
	plane for different values of $\Phi_0$ when $m_t=175$ GeV, $\tan{\beta}=2$ and $\mu<0$. 
	The allowed region lies below the curves.
	}
	\includegraphics[angle=270,width=3.1in]{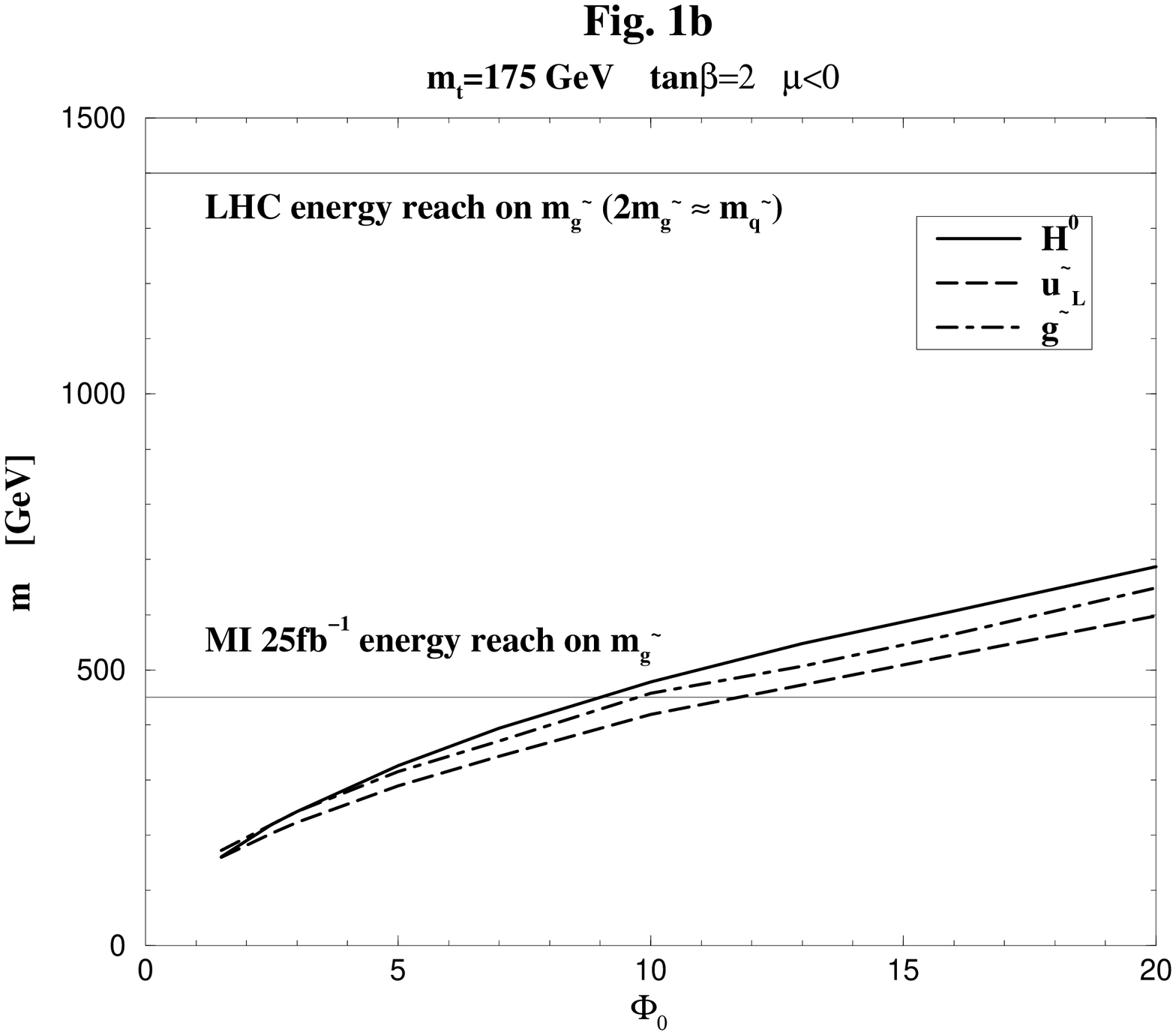} 
	\small{
	Fig.1b. Upper bounds on mass of the heavy Higgs $H^0$, of the gluino and of the squark 
	$\tilde u_L$ (for the first two generations) for the same parameters as in Fig.1a.
	}
\end{minipage}
\hfill
\begin{minipage}[t]{3.0in}
	\includegraphics[angle=270,width=3.1in]{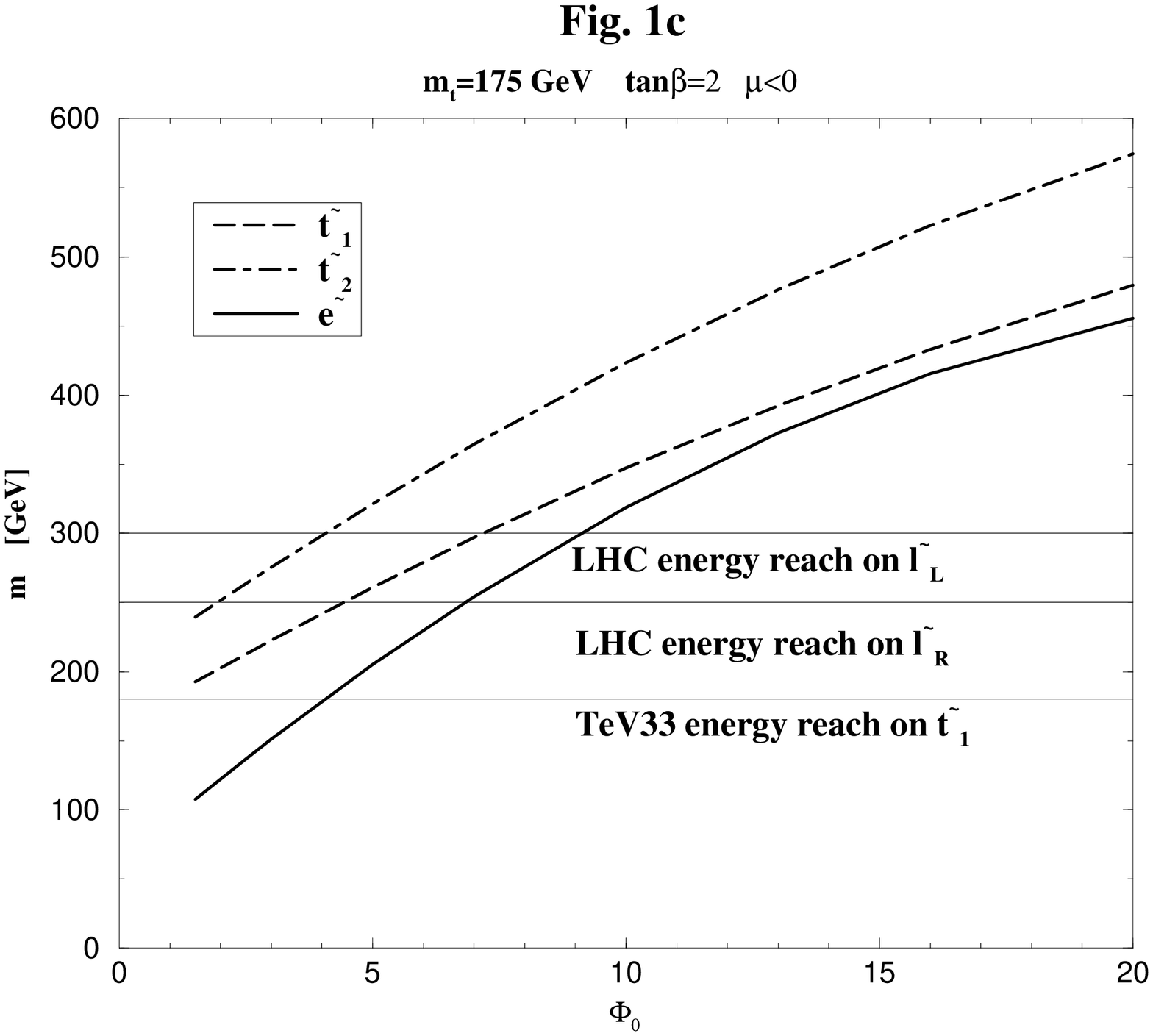} 
	\small{ 
	Fig.1c.  Upper bounds on mass of the $\tilde e_L$, of the light stop $\tilde t_1$, and of the 
	heavy stop $\tilde t_2$ for the same parameters as in Fig.1a.
	}
	\includegraphics[angle=270,width=3.1in]{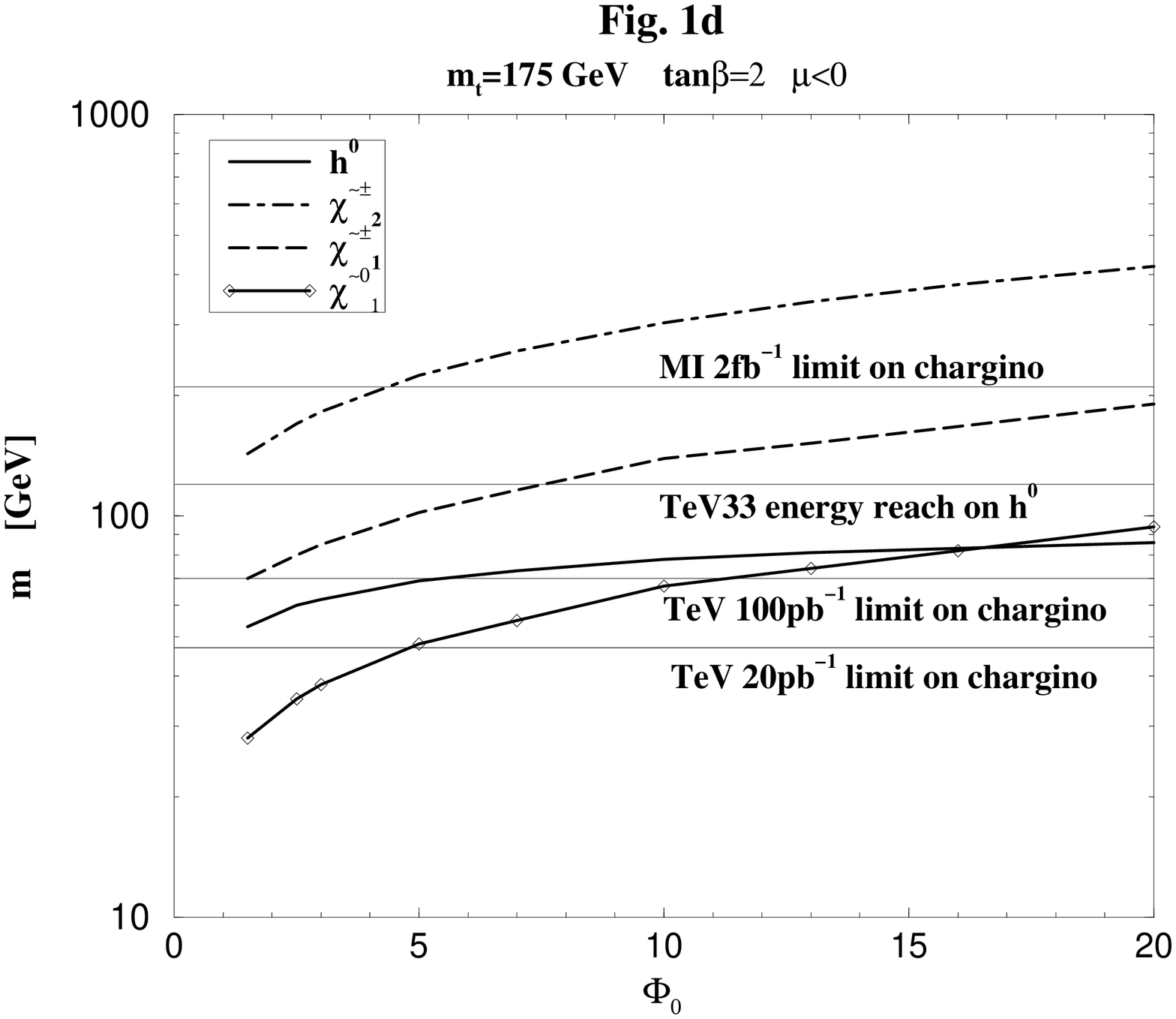} 
	\small{ 
	Fig.1d.  Upper bounds on masses of the light Higgs $h^0$, of the light chargino $\tilde \chi^\pm_1$,
	 of the heavy chargino $\tilde \chi^\pm_2$, and of the neutralino $\tilde \chi^0_1$ for the 
	 same parameters as in Fig.1a.
 	}
\end{minipage}
\end{figure}

\begin{figure}[htbp]
\begin{minipage}[t]{3.0in}
	\includegraphics[angle=270,width=3.1in]{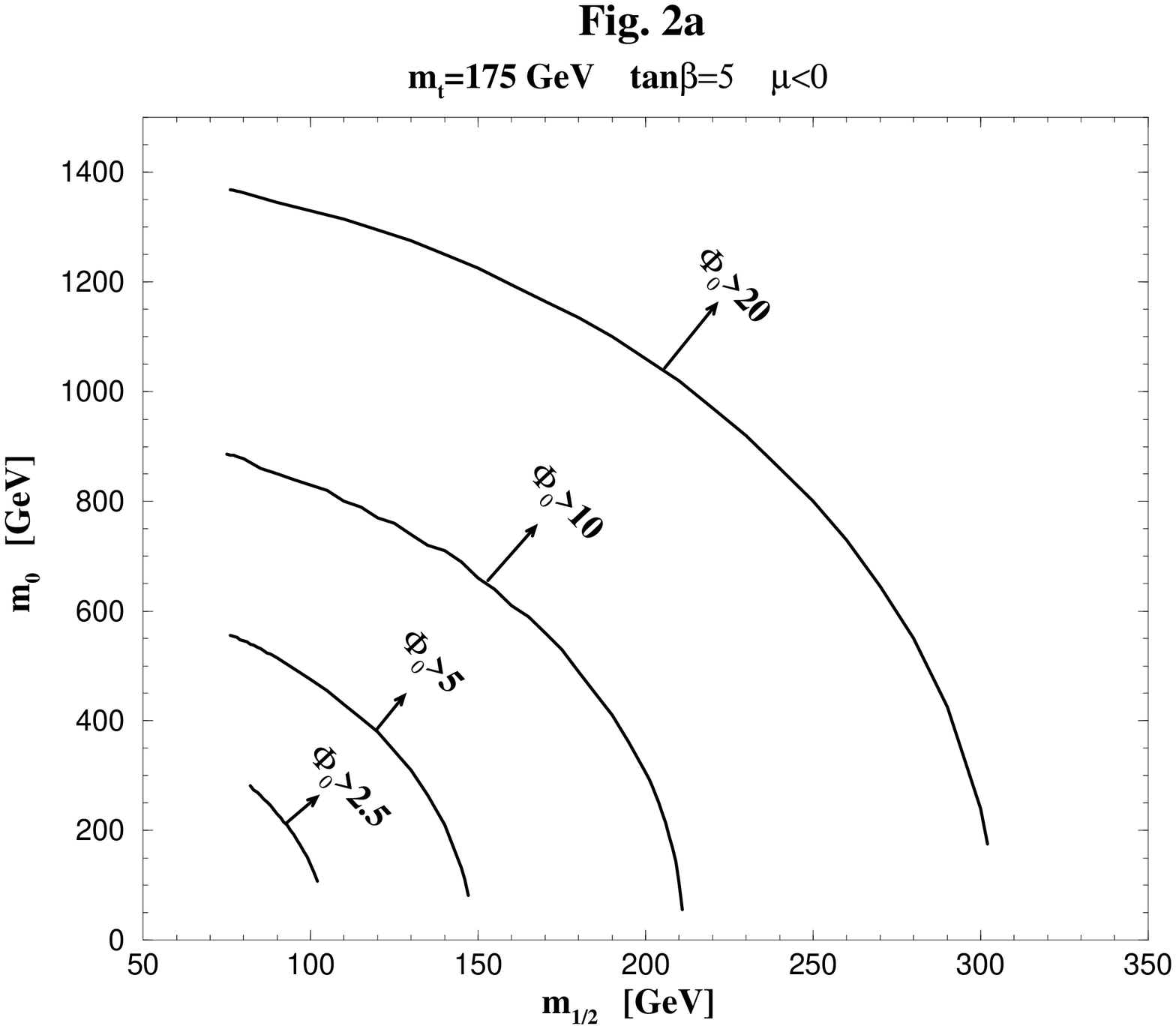}  
	\small{ 
	Fig.2a. Contour plot of the upper limit in the $m_0 - m_\frac{1}{2}$  
	plane for different values of $\Phi_0$ when $m_t=175$ GeV, $\tan{\beta}=5$ and $\mu<0$. 
	The allowed region lies below the curves.
 	}
	\includegraphics[angle=270,width=3.1in]{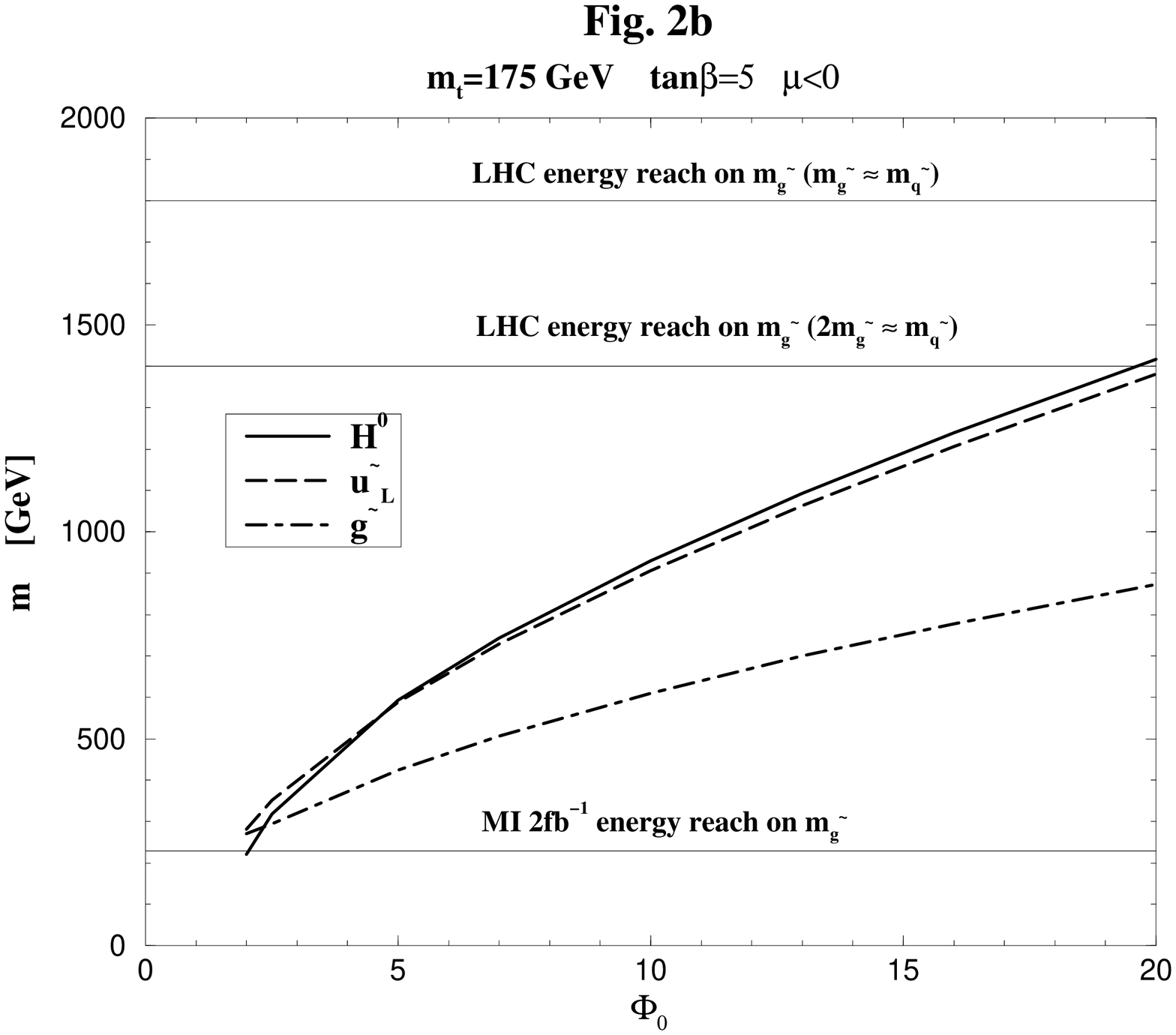}  
	\small{ 
	Fig.2b. Upper bounds on mass of the heavy Higgs $H^0$, of the gluino and of the squark 
	$\tilde u_L$ (for the first two generations) for the same parameters as in Fig.2a.
 	}
\end{minipage}
\hfill
\begin{minipage}[t]{3.0in}
	\includegraphics[angle=270,width=3.1in]{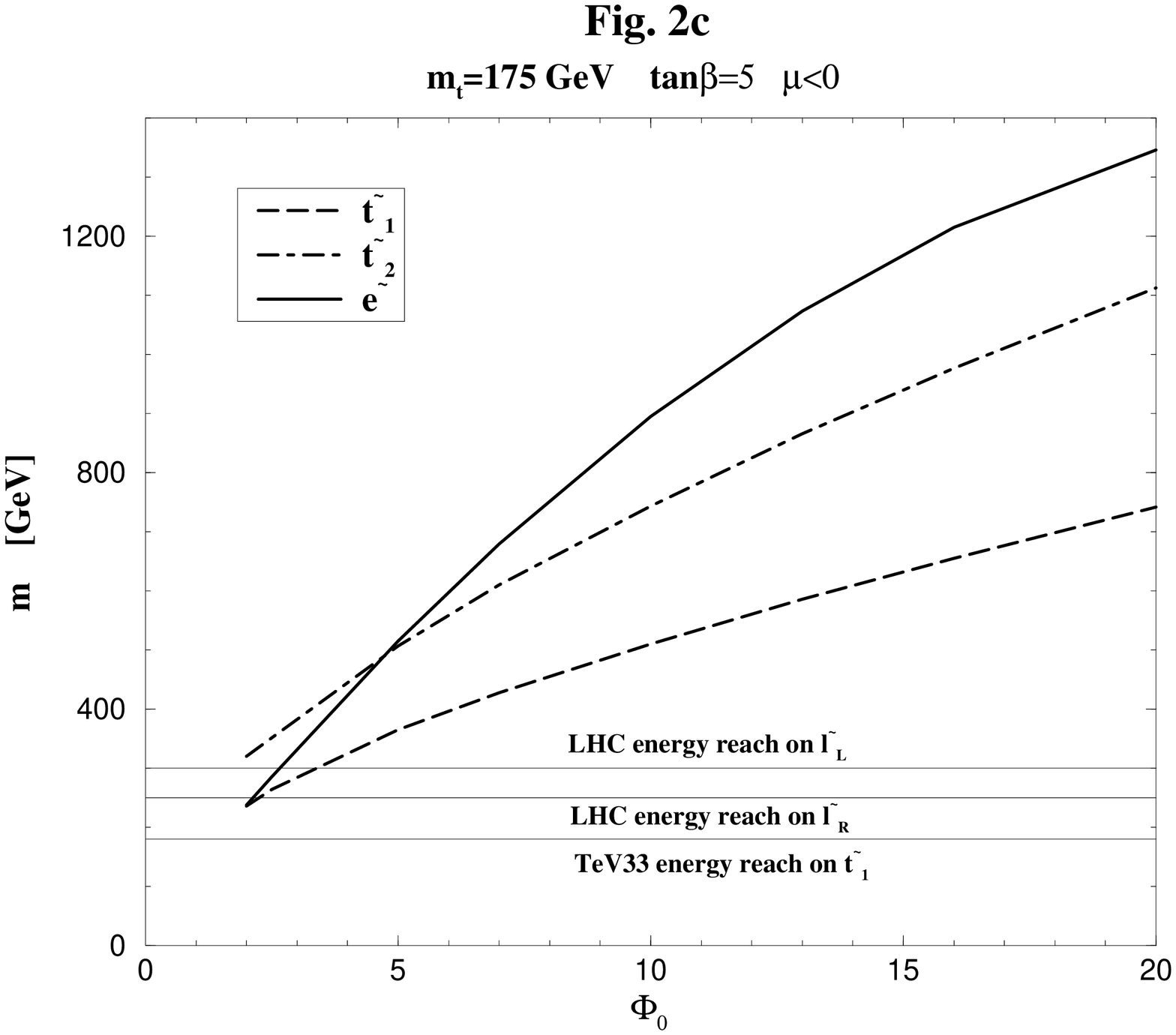}  
	\small{ 
	Fig.2c.  Upper bounds on mass of the $\tilde e_L$, of the light stop $\tilde t_1$, and of the 
	heavy stop $\tilde t_2$ for the same parameters as in Fig.2a.
 	}
	\includegraphics[angle=270,width=3.1in]{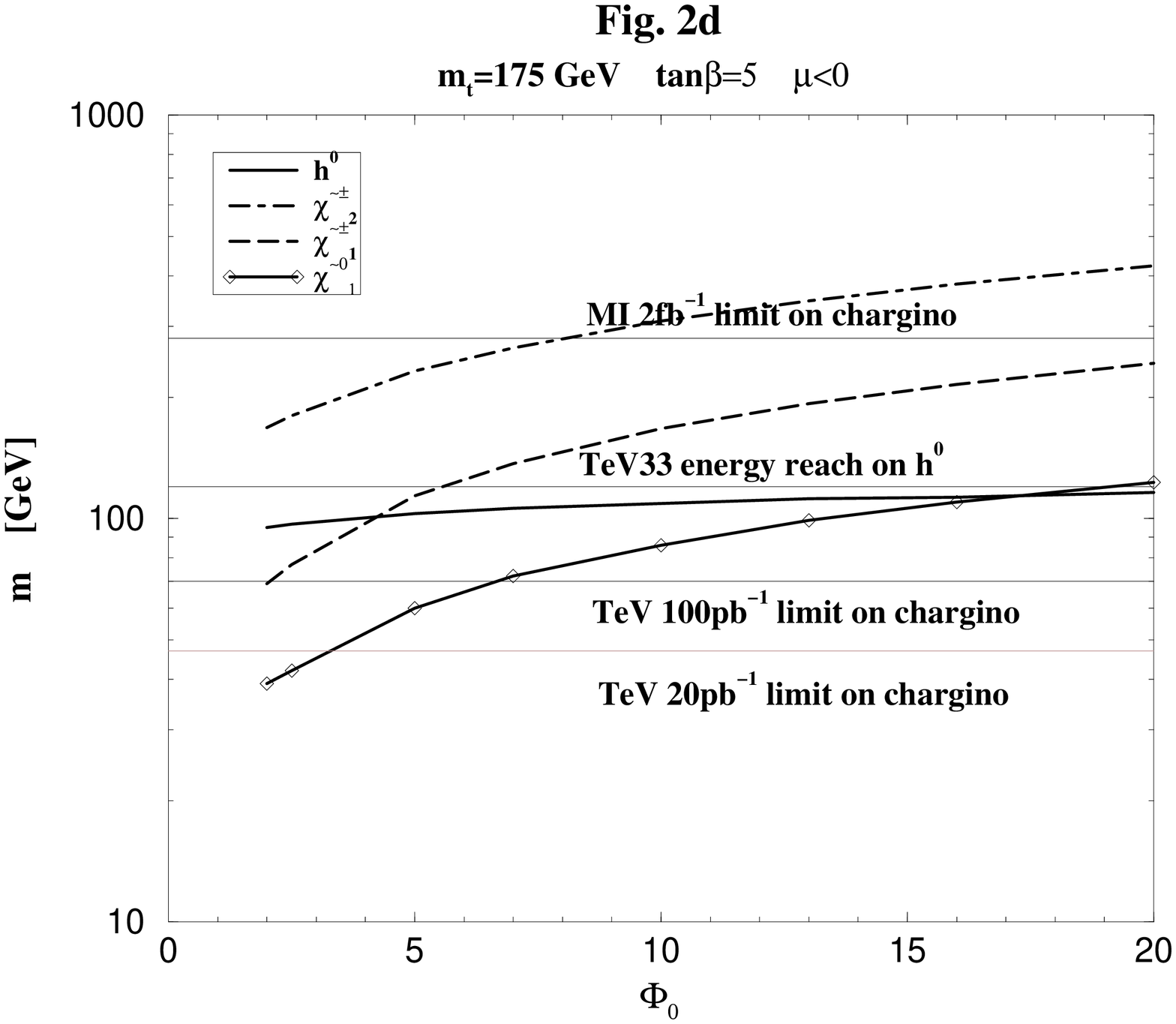}  
	\small{ 
	Fig.2d.  Upper bounds on masses of the light Higgs $h^0$, of the light chargino $\tilde \chi^\pm_1$,
	 of the heavy chargino $\tilde \chi^\pm_2$, and of the neutralino $\tilde \chi^0_1$ for the 
	 same parameters as in Fig.2a.
 	}
\end{minipage}
\end{figure}

\begin{figure}[htbp]
\begin{minipage}[t]{3.0in}
	\includegraphics[angle=270,width=3.1in]{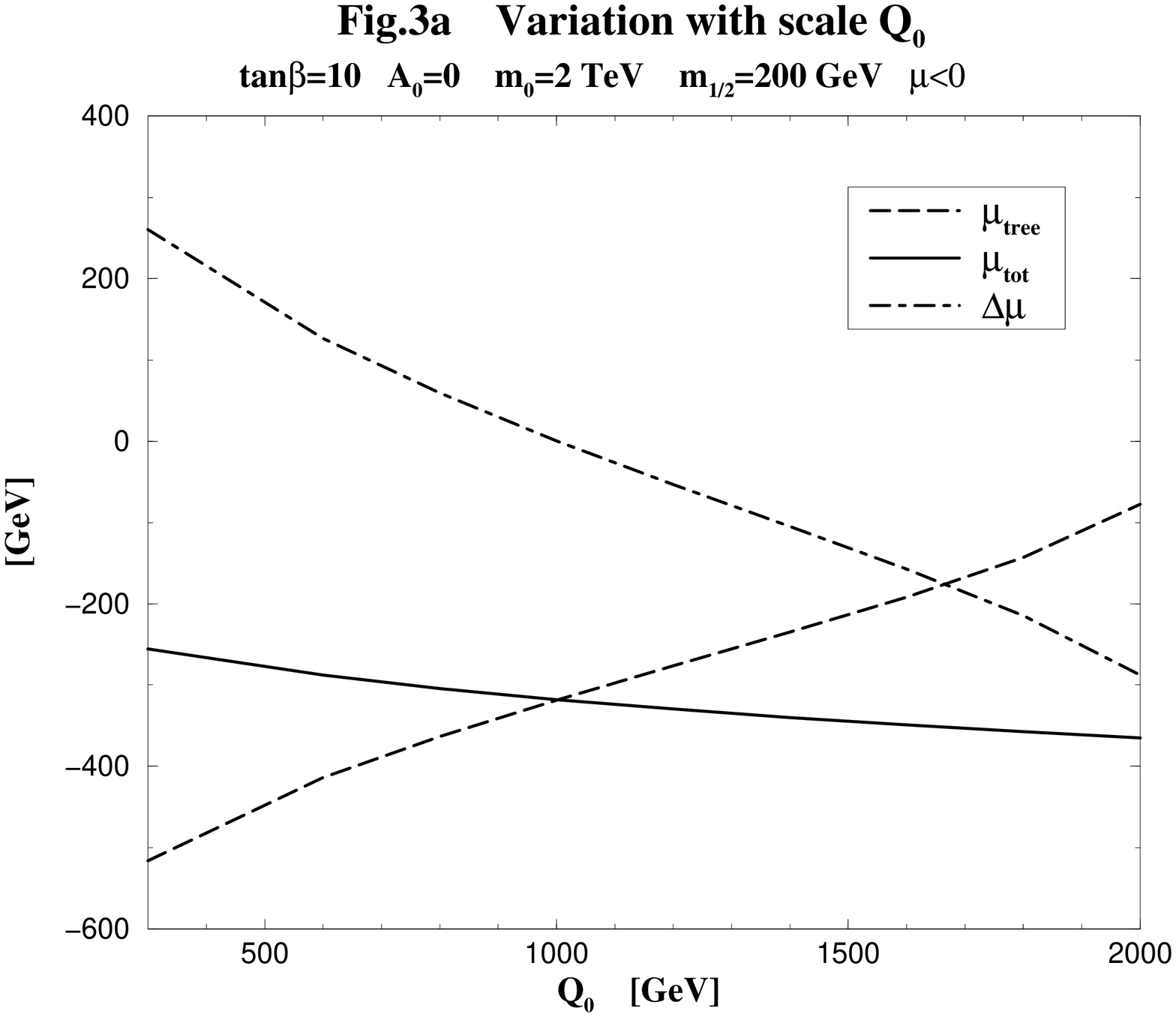}  
	\small{ 
	Fig.3a. Variation of $\mu$ with the scale $Q_0$ where the 
	minimization of the potential is carried out for
	the case when  $\tan{\beta}=10$,
	$A_0=0$, $m_0=2000$GeV, $m_{1/2}=$200 GeV and $\mu<0$.
 	}
\end{minipage}
\hfill
\begin{minipage}[t]{3.0in}
	\includegraphics[angle=270,width=3.1in]{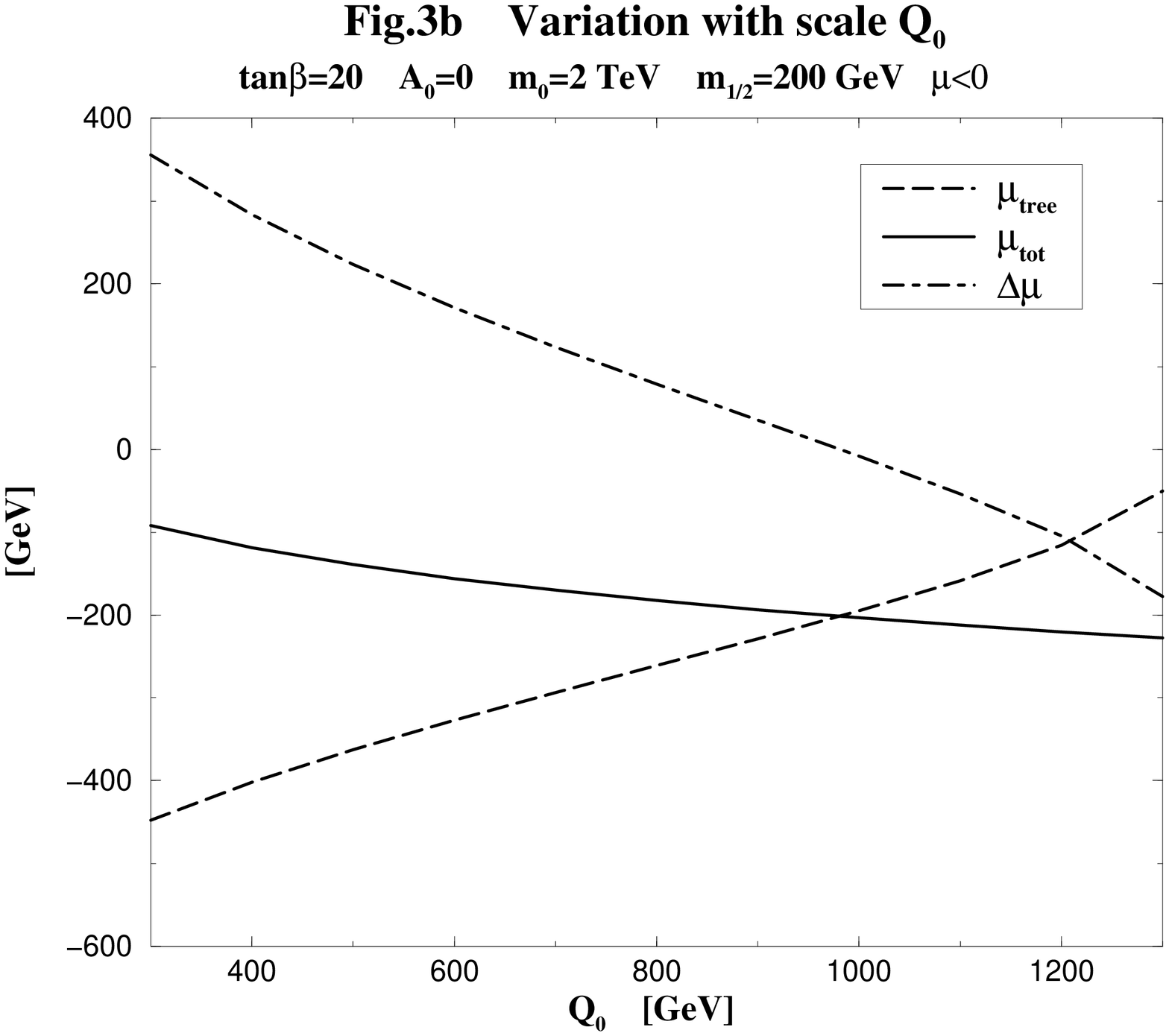}  
	\small{ 
	Fig.3b. Variation of $\mu$ with the scale $Q_0$ where the 
	minimization of the potential is carried out for the case 
	when $\tan{\beta}=20$,
	with the other parameters the same as in Fig.3a.
 	}
\end{minipage}
\end{figure}

\newpage

\begin{figure}[htbp]
\begin{minipage}[t]{3.0in}
	\includegraphics[angle=270,width=3.1in]{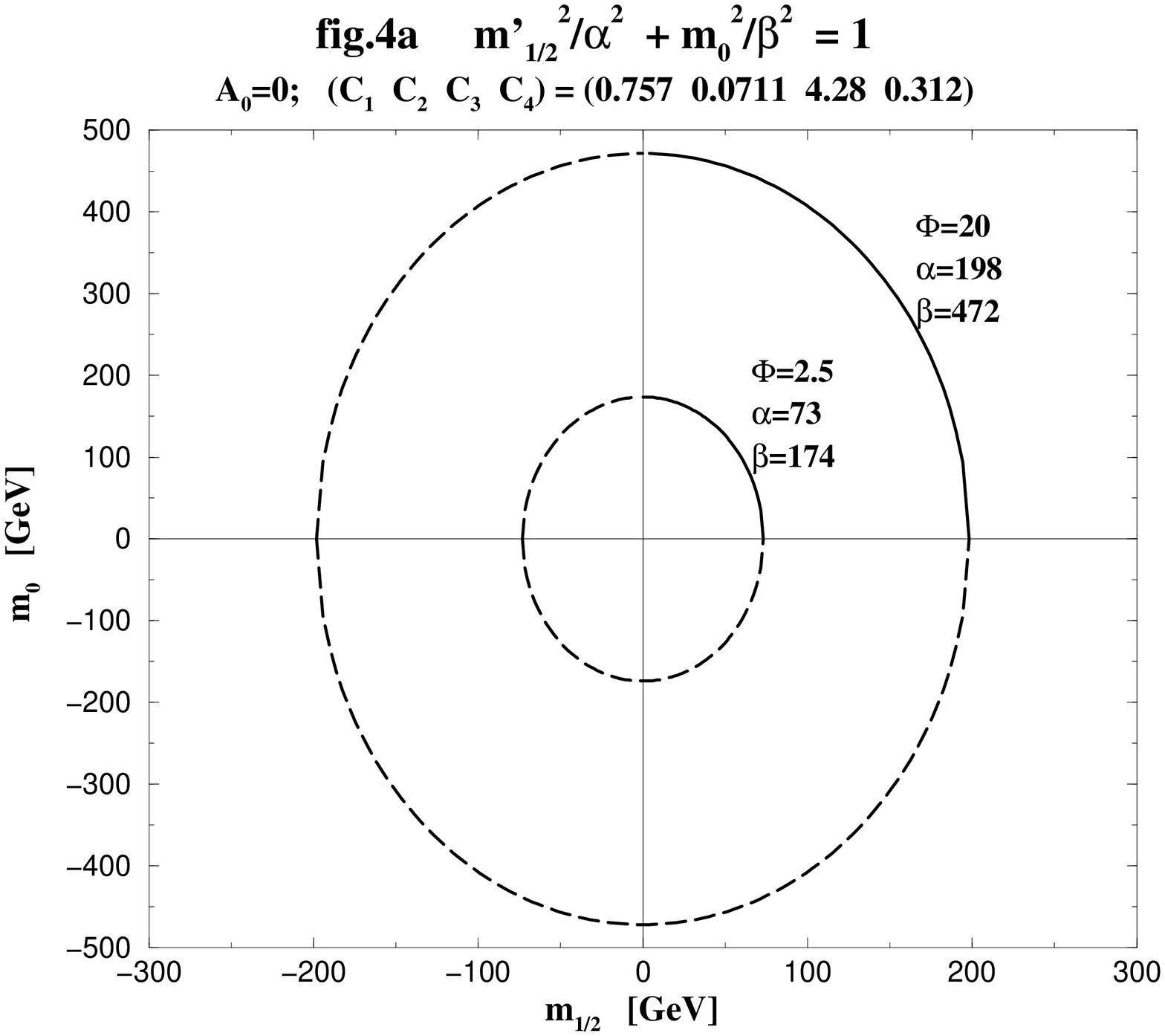}  
	\small{ 
	Fig.4a. Diagrammatic illustration of the ellipse represented by 
	Eq.(12), where the values of $C_1$ -- $C_4$ are for 
	tan$\beta$=2 and $Q=M_Z$  from Table 1. The relevant parts of the
	ellipses are in solid line.     
	 	}
	\includegraphics[angle=270,width=3.1in]{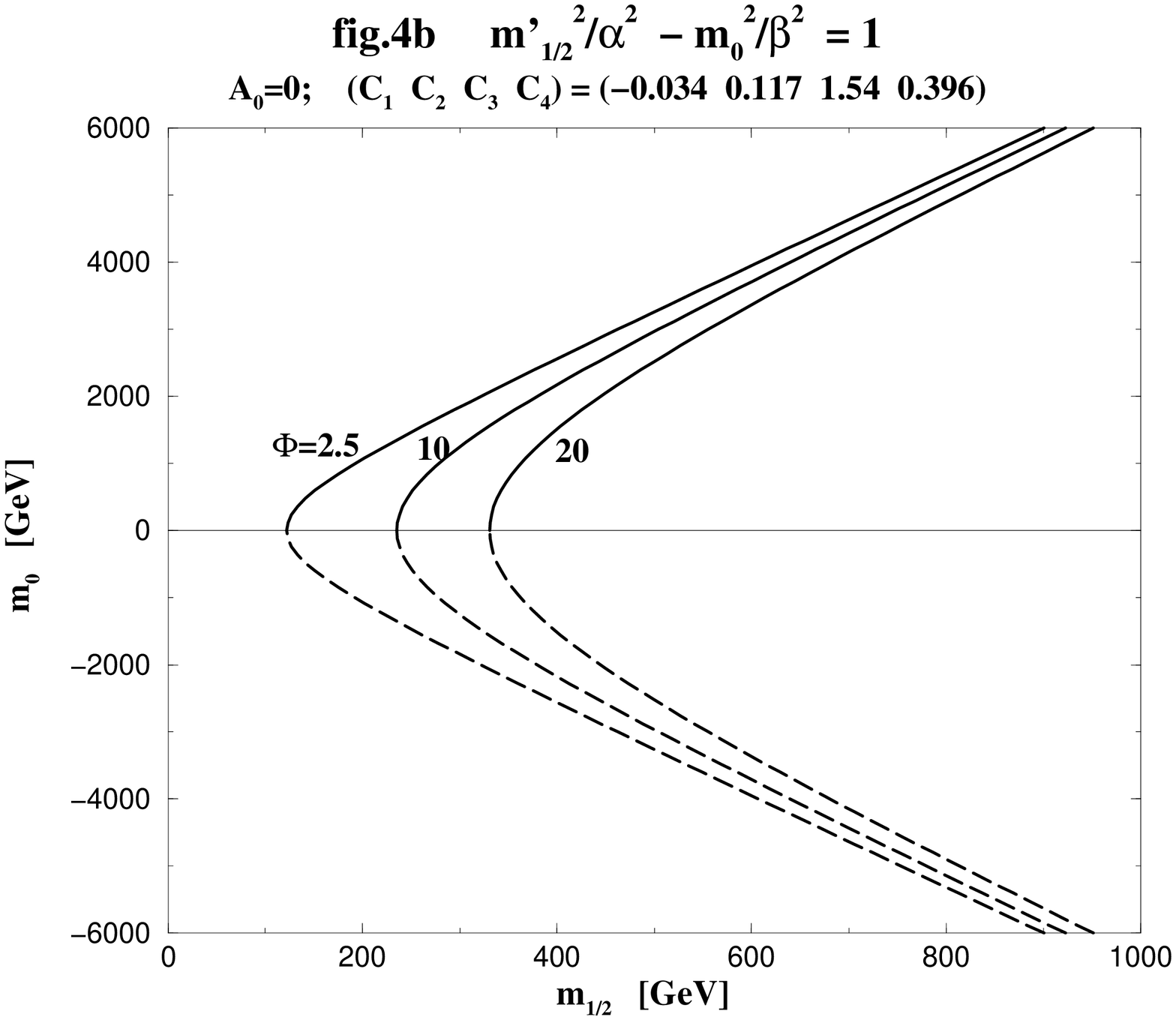}  
	\small{ 
	Fig.4b. Diagrammatic illustration of the hyperbola represented by 
	Eq.(14) and Eq.(28), where the values of $C_1$ -- $C_4$ are for 
	tan$\beta$=10 and Q=3000 GeV from Table 1.
	The relevant parts of the
	hyperbolae are in solid line.         
	 	}
\end{minipage}
\hfill
\begin{minipage}[t]{3.0in}
	\includegraphics[angle=270,width=3.1in]{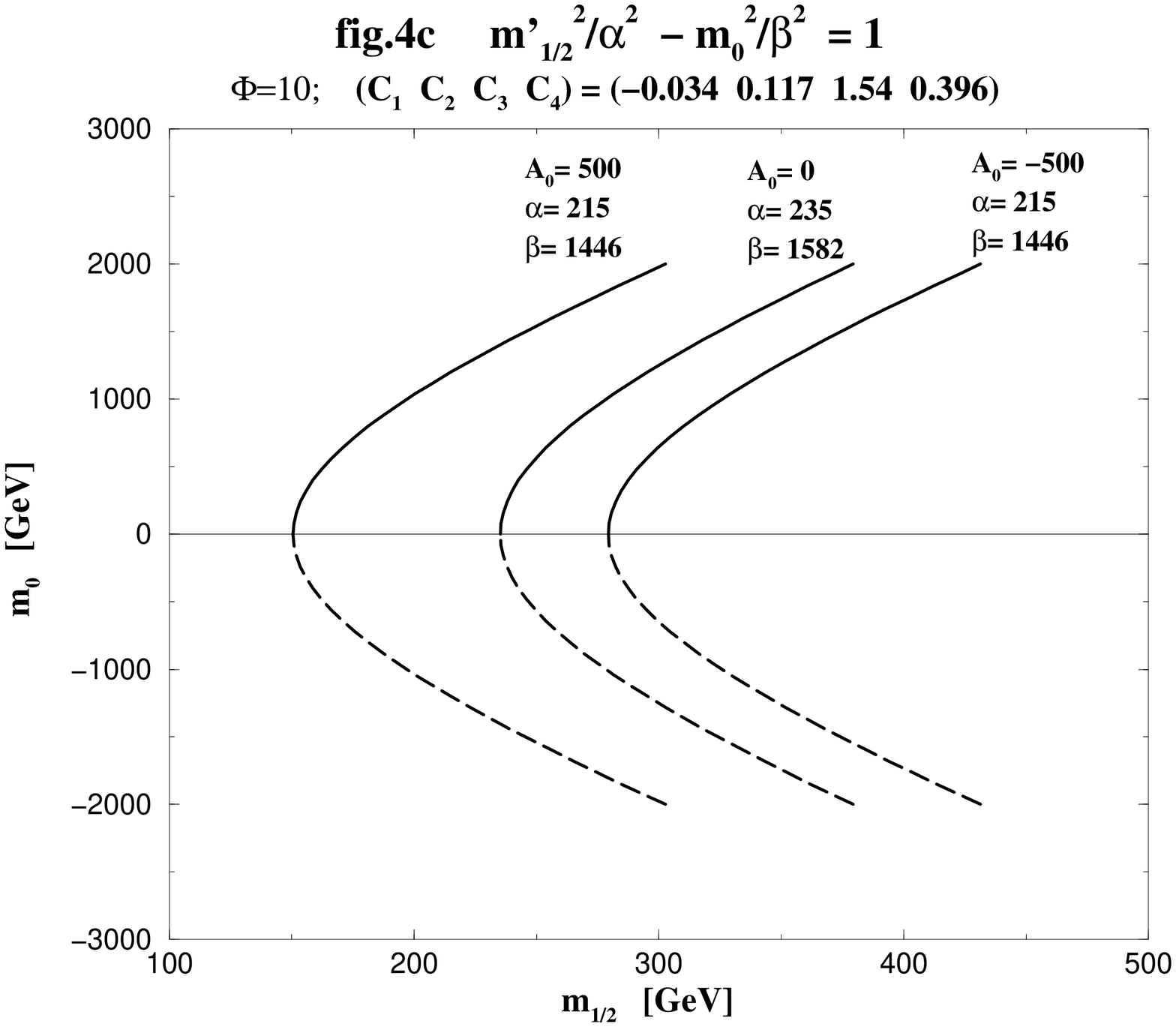}  
	\small{ 
	Fig.4c. Diagrammatic illustration of the hyperbola represented by 
	Eq.(14) and Eq.(28), where the values of $C_1$ -- $C_4$ are for 
	tan$\beta$=10 and Q=3000 GeV from Table 1.
	The relevant parts of the
	hyperbolae are in solid line.         
	 	}
	\includegraphics[angle=270,width=3.1in]{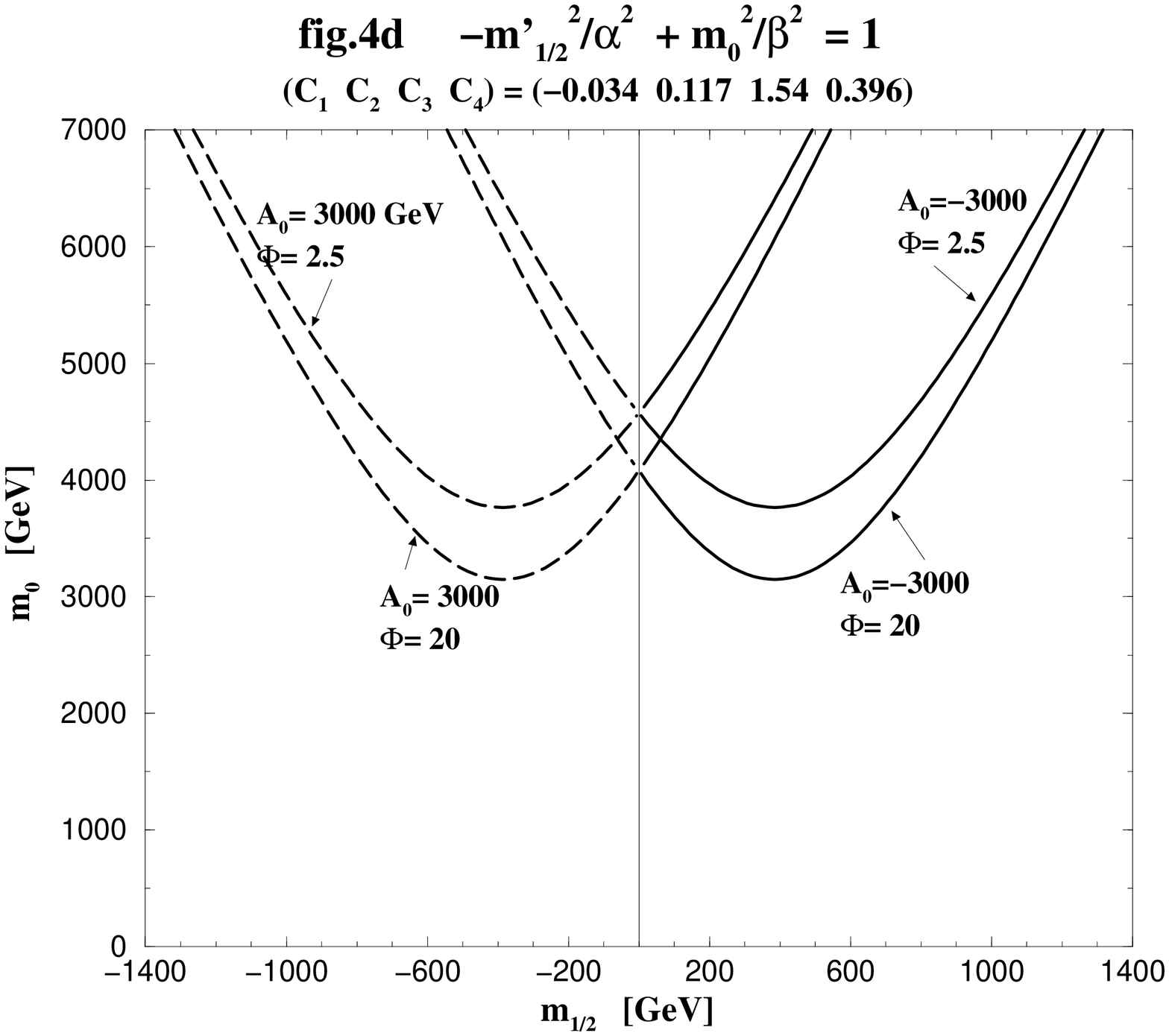}  
	\small{ 
	Fig.4d. Diagrammatic illustration of the hyperbola represented by 
	Eq.(19) and Eq.(30), where the values of $C_1$ -- $C_4$ are for 
	tan$\beta$=10 and Q=3000 GeV from Table 1.
	The relevant parts of the
	hyperbolae are in solid line.     	 	}
\end{minipage}
\end{figure}

\begin{figure}[htbp]			
\begin{minipage}[t]{6.0in}
	\includegraphics[angle=270,width=5.5in]{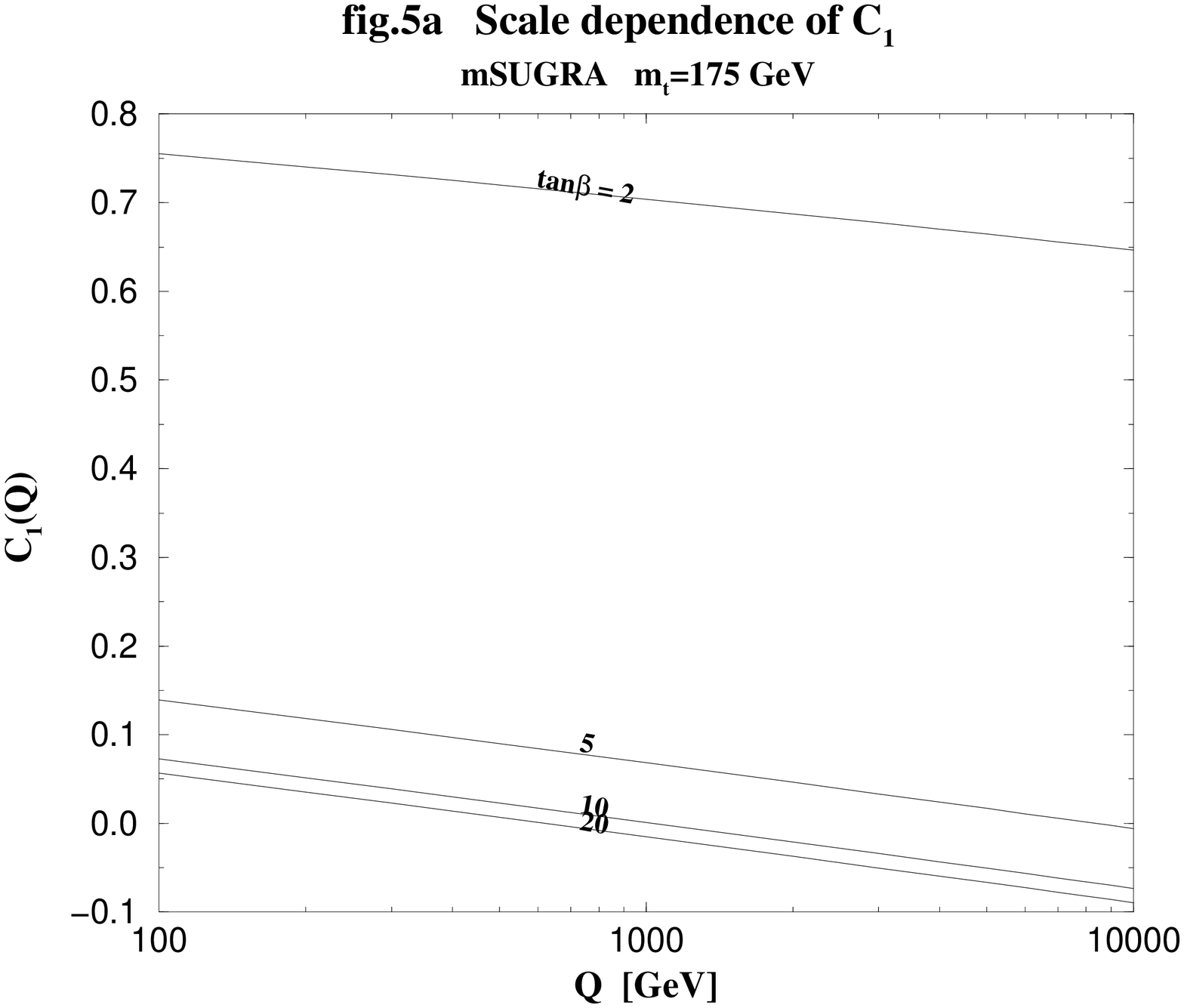}         
	\small{ 
	Fig.5a. The  scale dependence of $C_1(Q)$ 
	 for minimal 
	supergravity when  
	$m_t=175~GeV$
	 for  $tan{\beta}=$ 2, 5, 10 and 20.
 	}\label{fig_scale1a}
\end{minipage}
\end{figure}

\begin{figure}[htbp]
\begin{minipage}[t]{6.0in}
	\includegraphics[angle=270,width=6.0in]{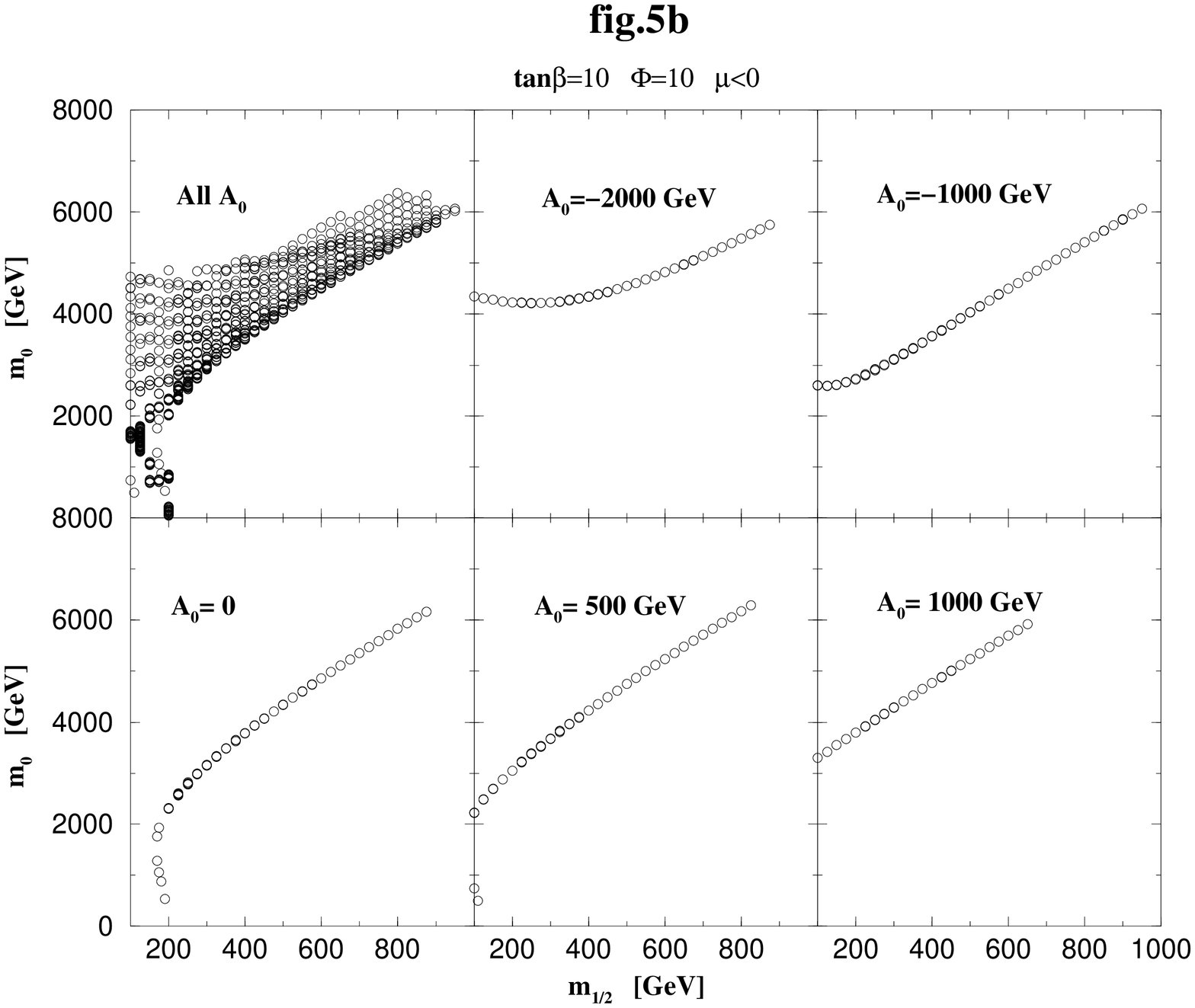}         
	\small{ 
	Fig.5b. Allowed region in the $m_0$ -- $m_{1/2}$ plane in 
	the minimal supergravity case for $m_t=175$ GeV,  
	tan$\beta$=10, $\Phi_0 =$10 
	and negative $\mu$. 
 	}
\end{minipage}
\end{figure}

\begin{figure}[htbp]			
\begin{minipage}[t]{6.0in}
	\includegraphics[angle=270,width=6.0in]{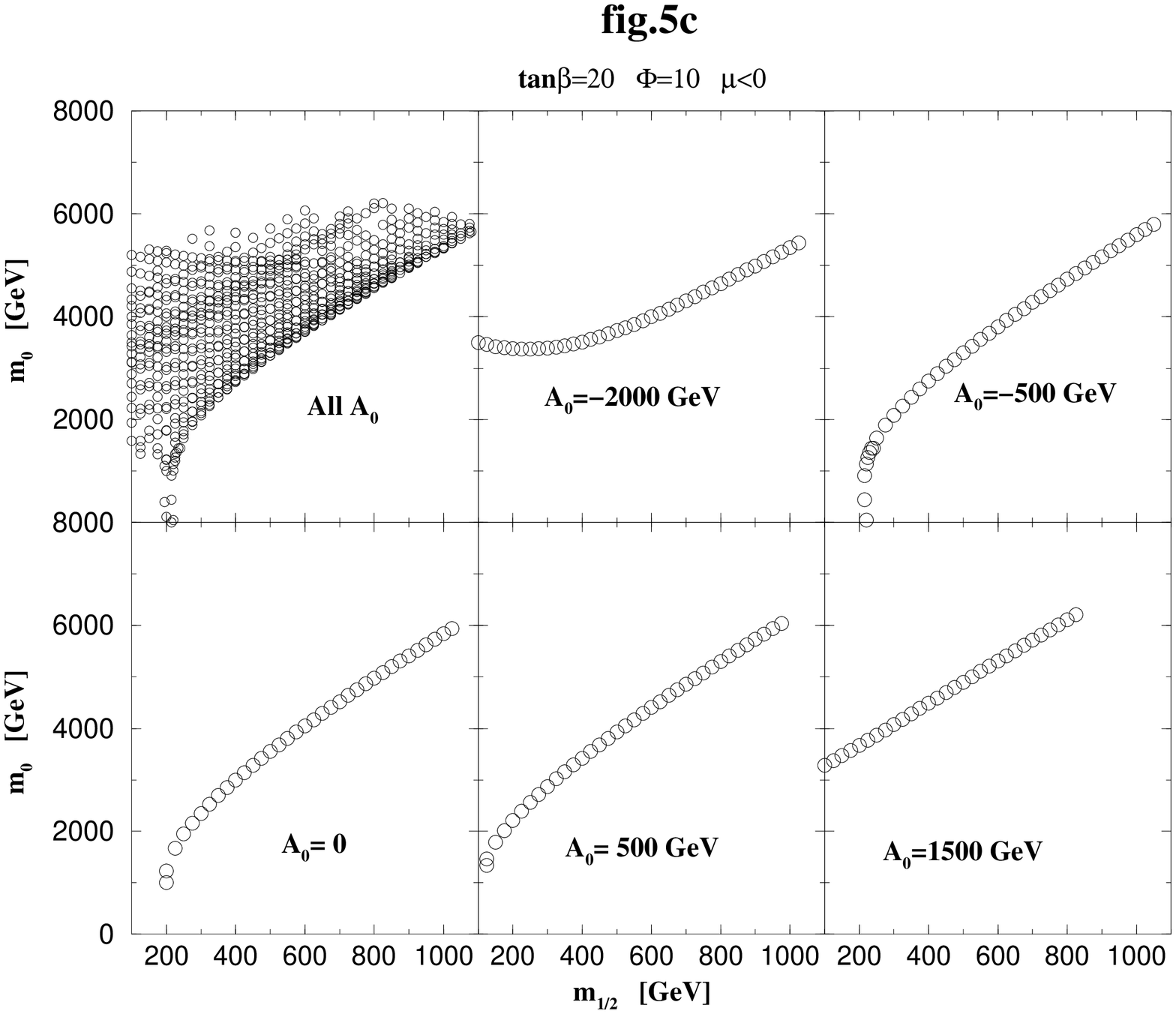}        
	\small{ 
	Fig.5c. Allowed region in the $m_0$ -- $m_{1/2}$ plane in 
	the minimal supergravity case for $m_t=175$ GeV,  
	tan$\beta$=20, $\Phi_0 =$10 
	and negative $\mu$. 
 	}
\end{minipage}
\end{figure}

\begin{figure}[htbp]
\begin{minipage}[t]{6.0in}
	\includegraphics[angle=270,width=6.0in]{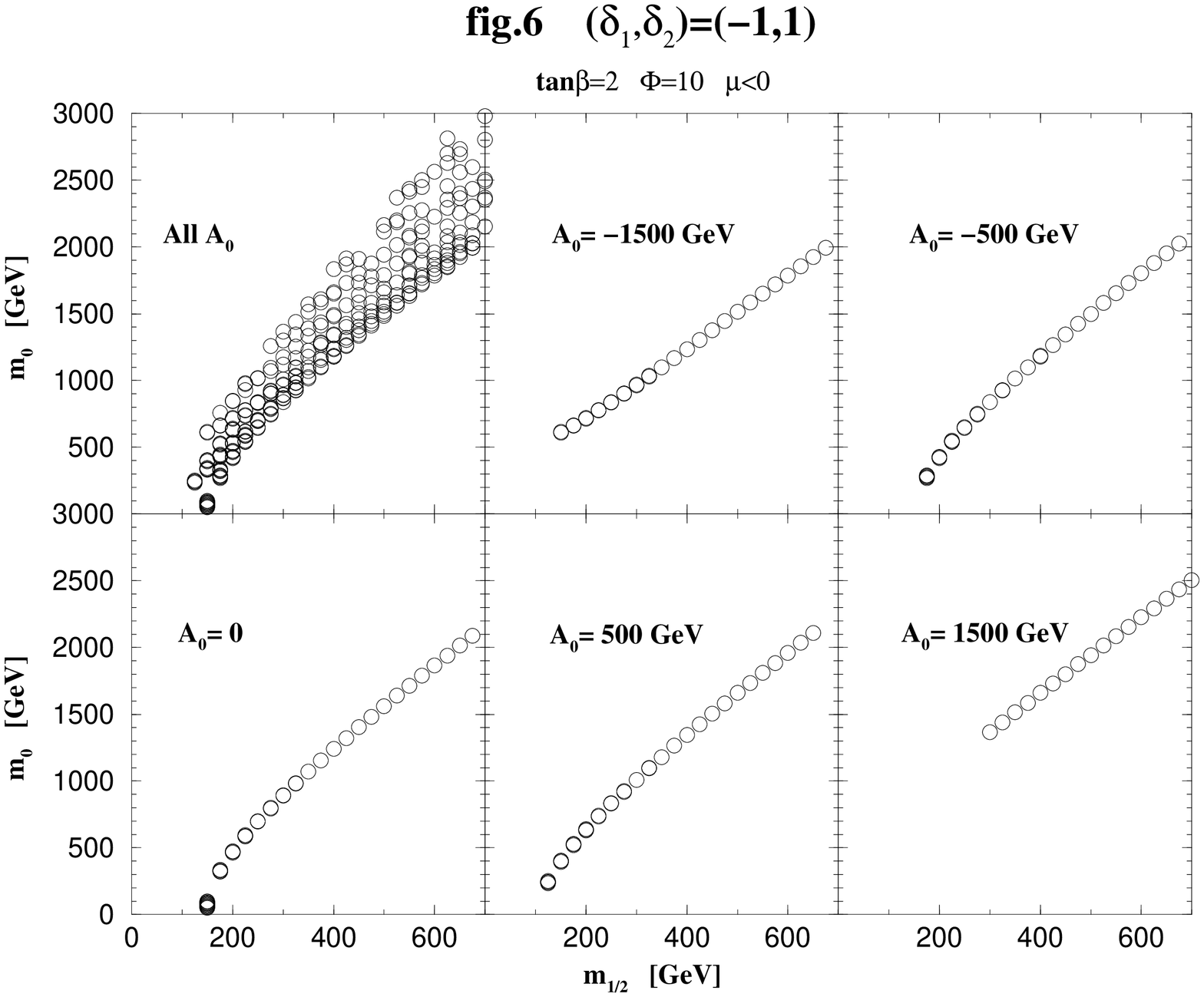}         
	\small{ 
	Fig.6. Allowed region in the $m_0$ -- $m_{1/2}$ plane  
	under the non-universal boundary condition of ($\delta_1,\delta_2$)=(-1,1)
	 for $m_t=175$ GeV,  
	tan$\beta$=2, $\Phi =$10 
	and negative $\mu$. 
 	}
\end{minipage}
\end{figure}

\begin{figure}[htbp]
\begin{minipage}[t]{6.0in}
	\includegraphics[angle=270,width=6.0in]{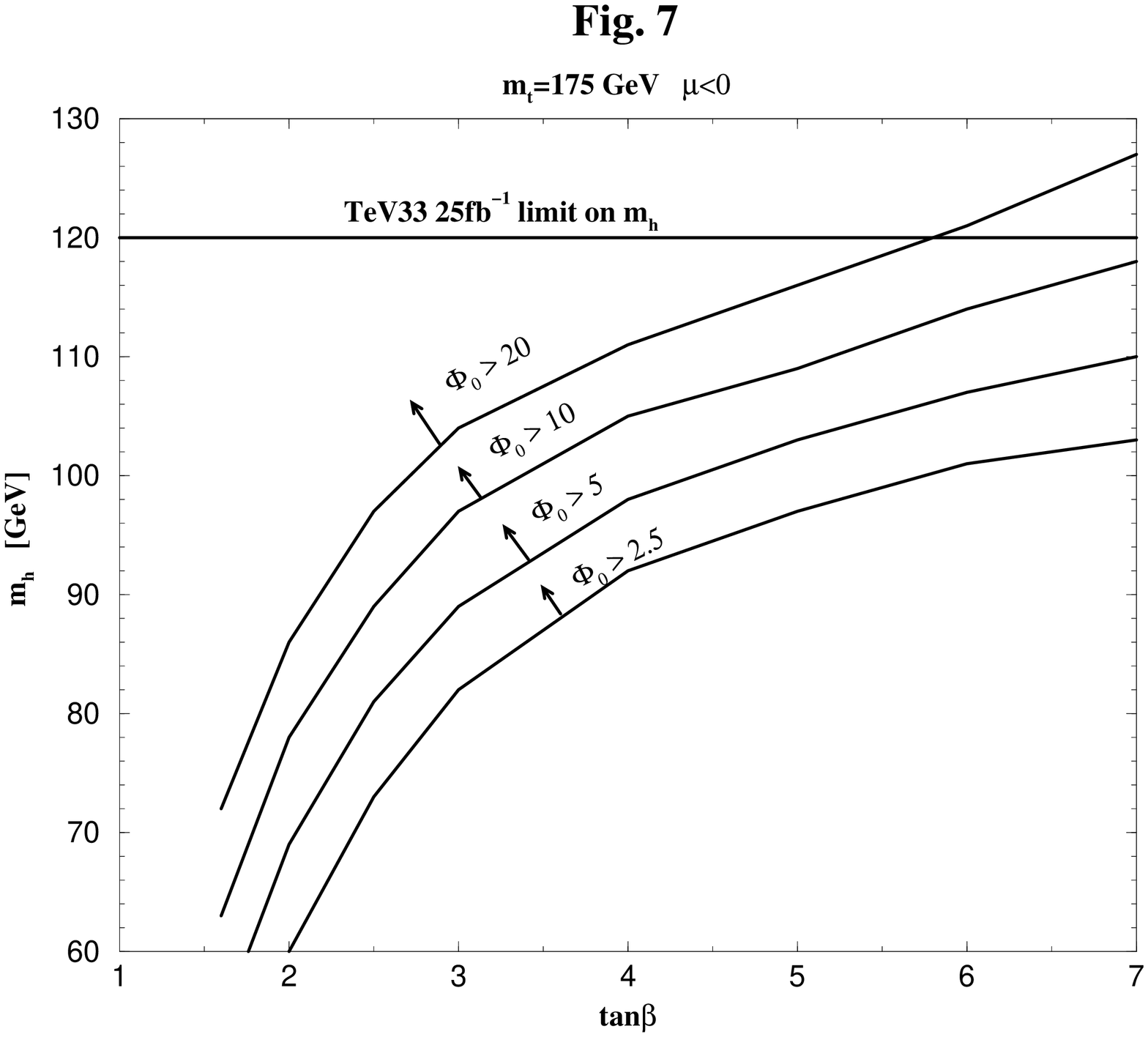}         
	\small{ 
	Fig.7. Upper bounds on the light Higgs $h^0$ mass  for
	different values of $\Phi_0$ as a function of $\tan{\beta}$ 
	when $m_t=175$ GeV and $\mu<0$.
 	}
\end{minipage}
\end{figure}

\end{document}